\documentclass[letter,11pt]{article}
\usepackage[top=1in,bottom=1in,left=1in,right=1in]{geometry}
\usepackage{slashed}
\usepackage{epsfig}
\usepackage{comment}
\usepackage{cancel}
\usepackage{bbm}
\usepackage{array}
\usepackage{bigints}
\usepackage{booktabs}
\usepackage{color}
\usepackage{dsfont}
\usepackage{float}
\usepackage{framed}
\usepackage{graphicx}
\usepackage{indentfirst}
\usepackage{mathrsfs}
\usepackage{multirow}
\usepackage{subdepth}
\usepackage{titlesec}
\usepackage[dotinlabels]{titletoc}
\usepackage{wrapfig}
\usepackage[all]{xy}
\usepackage[vcentermath]{youngtab}
\usepackage{relsize}
\usepackage{hyperref}
\numberwithin{equation}{section}

\usepackage[utf8]{inputenc}
\usepackage{slashed}
\usepackage{amsmath}
\usepackage{amsfonts}
\usepackage{amssymb}
\usepackage{cite}
\usepackage{mathtools}

\usepackage{cleveref}
\crefname{section}{§}{§§}
\Crefname{section}{§}{§§}

 \def\p{\partial}
 \def\bz{{\bar z}}
 \def\bw{{\bar w}}
  \def\bh{{\bar h}}
 
\def\0{{(0)}}
\def\1{{(1)}}
\def\2{{(2)}} 
  
\def\co{{\cal O}}

\def\<{\langle }
\def\>{\rangle }
\def\bw{{\bar w}}

\newcommand{\bea}{\begin{eqnarray}}
\newcommand{\eea}{\end{eqnarray}}
\newcommand{\be}{\begin{equation}}
\newcommand{\ee}{\end{equation}}
\newcommand{\ba}{\begin{align}}
\newcommand{\ea}{\end{align}}

   \makeatletter
  \let\over=\@@over \let\overwithdelims=\@@overwithdelims
  \let\atop=\@@atop \let\atopwithdelims=\@@atopwithdelims
  \let\above=\@@above \let\abovewithdelims=\@@abovewithdelims
\renewcommand\section{\@startsection {section}{1}{\z@}%
                                   {-3.5ex \@plus -1ex \@minus -.2ex}
                                   {2.3ex \@plus.2ex}%
                                   {\normalfont\large\bfseries}}

\renewcommand\subsection{\@startsection{subsection}{2}{\z@}%
                                     {-3.25ex\@plus -1ex \@minus -.2ex}%
                                     {1.5ex \@plus .2ex}%
                                     {\normalfont\bfseries}}

\newcommand{\beq}{\begin{equation}}
\newcommand{\eeq}{\end{equation}}
\newcommand{\beqa}{\begin{eqnarray}}
\newcommand{\eeqa}{\end{eqnarray}}
\newcommand{\beqar}{\begin{eqnarray*}}

\def\[{\big[}
\def\]{\big]}

\def\bz{{\bar z}}
\def\ba{{\bar a}}

\def\bh{{\bar h}}

\def\be{{\bar \epsilon}}

\def\bw{{\bar w}}

\def\eps{\epsilon}

\def\+{{(+)}}
\def\-{{(-)}}
\def\0{{(0)}}
\def\1{{(1)}}
\def\2{{(2)}}
\def\3{{(3)}}

\def\be{\begin{equation}}
\def\ee{\end{equation}}

\begin{document}
\begin{titlepage}
\unitlength = 1mm
\ \\
\vskip 3cm
\begin{center}
 
{\huge{\textsc{Celestial Operator Product Expansions
\vskip 1ex and ${\rm w}_{1+\infty}$ Symmetry for All Spins}}}

\vspace{1.25cm}
Elizabeth Himwich,$^{*\mathsection}$ Monica Pate,$^{*\mathsection \dagger}$ and Kyle Singh$^\ddagger$

\vspace{.5cm}

$^*${\it  Center for the Fundamental Laws of Nature, Harvard University,
Cambridge, MA 02138}\\ 
$^\mathsection${\it Black Hole Initiative, Harvard University, Cambridge, MA 02138}
\\ 
$^\dagger${\it  Society of Fellows, Harvard University,
Cambridge, MA 02138}\\ 
$^\ddagger${\it  Department of Physics, University of Pennsylvania, Philadelphia, PA 19104}

\vspace{0.8cm}

\begin{abstract} 

The operator product expansion of massless celestial primary operators of arbitrary spin is investigated.  Poincar\'e symmetry is found to imply a set of recursion relations on the operator product expansion coefficients of the leading singular terms at tree-level in a holomorphic limit. The symmetry constraints are solved by an Euler beta function with arguments that depend simply on the right-moving conformal weights of the operators in the product.  These symmetry-derived coefficients are shown not only to match precisely those arising from momentum-space  tree-level collinear limits, but also to obey an infinite number of additional symmetry transformations that respect the algebra of ${\rm w}_{1+\infty}$.  In tree-level minimally-coupled gravitational theories, celestial currents are constructed from light transforms of conformally soft gravitons and found to generate the action of ${\rm w}_{1+\infty}$ on arbitrary massless celestial primaries. Results include operator product expansion coefficients for fermions as well as those arising from  higher-derivative non-minimal couplings of gluons and gravitons.

\end{abstract}

\vspace{1.0cm}

\end{center}

\end{titlepage}

\pagestyle{empty}
\pagestyle{plain}

\def\vx{{\vec x}}
\def\p{\partial}
\def\po{$\cal P_O$}
\def\i{{\rm initial}}
\def\f{{\rm final}}

\pagenumbering{arabic}
 

\tableofcontents

\section{Introduction}

	The holographic principle, which purports that a quantum theory of gravity can be captured by an ordinary non-gravitational theory in fewer dimensions, offers a profound perspective on the nature of quantum gravity.   
	At present, the AdS/CFT correspondence \cite{Maldacena:1997re} remains our best-established and most thoroughly investigated example of this paradigm.  In asymptotically flat spacetimes, the scattering problem nevertheless suggests a way in which the holographic principle might naturally extend to this context. In particular, since the observables associated with scattering in asymptotically flat space depend only on data characterizing the states at asymptotically early and late times,  they reside by construction in the fewer dimensions of the past and future boundaries of spacetime.

	This simple reasoning is further supported and clarified by the observation that the Lorentz symmetry ${\rm SO}(3, 1)$ of scattering in four dimensions  is isomorphic to the global conformal symmetry ${\rm SL}(2, \mathbb{C})$ of theories in two dimensions.  In other words, scattering in four dimensions is organized according to the global conformal symmetry of a theory in \emph{two} dimensions! Therefore, a theory in two dimensions that admits a conformal symmetry is a natural candidate for the holographic dual of quantum gravity in four-dimensional asymptotically flat spacetime.  
	 
	 A more thorough investigation of this proposal is aided by working in a language in which the underlying two dimensions are rendered manifest. 2D conformal (4D Lorentz) symmetry is the basis of this two-dimensional description, so   these two dimensions are more readily apparent in the scattering of particles of definite boost weight, as opposed to the standard  momentum eigenstates.  More precisely, one can construct boost-eigenstate scattering amplitudes that transform under Lorentz  transformations like correlation functions of primary operators under global conformal transformations \cite{Pasterski:2016qvg,Pasterski:2017kqt,Pasterski:2017ylz}. These amplitudes are referred to as  celestial amplitudes. 
	 
	 It is crucial for the existence of  an intrinsically-defined holographic dual theory that scattering amplitudes not only exhibit a global conformal symmetry, but also admit other  behavior characteristic of a two-dimensional theory. The Ward identities for infinite-dimensional symmetries that follow from soft theorems are powerful and encouraging examples of this behavior.    In particular, the subleading soft graviton theorem implies that the global conformal symmetry is enhanced to a local Virasoro symmetry generated by a stress tensor  \cite{Kapec:2014opa,Kapec:2016jld,Cheung:2016iub,Fotopoulos:2019tpe}.  Similarly, the leading soft theorem in gauge theory implies a  local Kac-Moody symmetry generated by a Kac-Moody current \cite{Strominger:2013lka,He:2015zea,Nande:2017dba,Fan:2019emx,Pate:2019mfs,Nandan:2019jas}.  Finally, soft theorems at other orders in gauge and gravitational  theories give rise to additional infinite-dimensional symmetries generated by generalized  2D currents \cite{Donnay:2018neh,Himwich:2019dug,Adamo:2019ipt,Puhm:2019zbl,Guevara:2019ypd,Fotopoulos:2019vac,Donnay:2020guq,Himwich:2020rro,Fotopoulos:2020bqj,Banerjee:2020zlg,Banerjee:2020vnt,Pasterski:2020pdk,Guevara:2021abz,Banerjee:2021cly, Pasterski:2021fjn,Pasterski:2021dqe,Jiang:2021xzy}.  This collection of currents was recently shown to admit a rich symmetry algebra \cite{Guevara:2021abz}, which has been identified as a ${\rm w}_{1 +\infty}$ algebra \cite{Strominger:2021lvk}.
	 
	  Remarkably, locality\footnote{Here we are referring to locality on the \emph{boundary}.  The implications of bulk locality were recently explored in \cite{Chang:2021wvv}.} -- in the form of an operator product expansion  (OPE) -- is yet another inherently two-dimensional property that 
	  has been established for  a class of boost eigenstates of massless particles \cite{Fan:2019emx}.   
	   These  2D-local OPEs originate from a collinear limit of momentum-space scattering amplitudes.  As a result, the leading terms in the OPE inherit the same universality that is associated to collinear singularities in momentum space.  For instance, the collinear splitting of graviton amplitudes is universal and does not receive loop corrections \cite{Bern:1998sv}, while gauge theory amplitudes admit collinear factorization in which the splitting function is corrected at every loop order (see e.g. \cite{Kosower:1999xi,Feige:2014wja}).

	  Taken together, the infinite-dimensional symmetries from soft theorems and the OPE offer new insight into constraints on the organization of scattering data of massless particles  \cite{Pate:2019lpp,Banerjee:2020kaa,Banerjee:2020zlg,Banerjee:2020vnt,Ebert:2020nqf,Banerjee:2021cly,Banerjee:2021dlm}. Notably, they form the basis of an approach to the scattering problem in which the symmetries arising from soft theorems are found to constrain interactions that do not involve any soft particles.  Massless particles -- irreducible representations of the Poincar\'e group -- are represented by a \textit{family} of operators of fixed spin and varying boost weight (conformal dimension) $\Delta$. In an OPE involving operators of fixed boost weight, the grouping of weights into single-particle families is captured by a non-trivial dependence of the OPE coefficients on the conformal dimensions $\Delta$ of the primaries. Standard conformal field theory analysis fixes the OPE up to the coefficients of primary operators, while 4D translations and the symmetries associated to soft theorems relate different conformal families, thereby imposing further constraints on the OPE coefficients.  

	  More specifically, in tree-level Einstein-Yang Mills theory, soft gluon and graviton theorems were shown to  supply enough structure to fix entirely the leading terms in gluon and graviton OPEs in a holomorphic limit \cite{Pate:2019lpp}. In this paper, we revisit the analysis of \cite{Pate:2019lpp} to determine OPE coefficients between massless particles of generic spin.  We demonstrate that 4D translations together with 4D Lorentz (2D global conformal) symmetry alone completely determine the $\Delta$-dependence of OPE coefficients between celestial operators of arbitrary spin at leading order in a holomorphic limit at tree-level.   The undetermined spin dependence of the OPE coefficients appears in the overall normalization and is effectively equivalent to the three-point coupling constant in a 4D effective action.  We explicitly compare all symmetry-derived OPE coefficients against those derived from tree-level collinear limits of momentum-space scattering amplitudes and find precise agreement.  
	  
	  In \cite{Pate:2019lpp} the subleading soft gluon theorem played an essential role in determining the gluon OPE coefficients, while the subsubleading soft graviton theorem played the analogous role for graviton OPE coefficients. Upon including minimal coupling between gluons and gravitons, the subleading  soft gluon theorem receives corrections \cite{Elvang:2016qvq,Laddha:2017vfh} that were critical for an appropriate generalization of the previous gluon OPE analysis. Our work shows that the tree-level coefficients of all gluon and graviton OPEs are fixed up to the 4D coupling constant by Poincar\'e symmetry alone, which is simply uncorrected and universal.  A similar Poincar\'e-based analysis for gluons in Yang-Mills theory was presented  in \cite{Ebert:2020nqf}.  

    Given that, in momentum space, it is well known that Poincar\'e symmetry uniquely fixes the three-point function between massless particles up to an overall coupling coefficient, our findings are not entirely surprising.  These three-point functions are directly related to the tree-level collinear splitting functions.  Poincar\'e symmetry is therefore sufficient at tree-level to fix leading celestial OPE coefficients  simply because these coefficients are the image of the splitting functions under a transformation to a boost weight basis.  Nevertheless, a precise implementation of this argument was  hitherto muddled, in part due to the singular nature of the three-point functions in boost-weight space \cite{Pasterski:2017ylz}.\footnote{ Boost-eigenstates of massless particles constructed with an additional light transformation were recently shown to admit smooth three-point functions,  which are more directly related to the OPE coefficients \cite{Sharma:2021gcz}.} 
	 
	 Moreover, for the purpose of constructing a holographically dual \emph{theory}, it is of substantial value to provide an intrinsically two-dimensional derivation of this result.  This is achieved by our analysis, in which ordinary 4D translation symmetry is regarded as an exotic global symmetry of the 2D theory.  By treating the symmetry generators in a manifestly 2D conformally covariant language, we readily determine their action on global conformal primaries and descendants, and thereby obtain constraints on OPE coefficients via standard CFT techniques.  In addition, we determine the OPE coefficients of descendants, for which we provide a closed-form  formula.\footnote{Previous investigations into asymptotic symmetry constraints on descendant OPE coefficients can be found in \cite{Banerjee:2020kaa,Ebert:2020nqf,Banerjee:2020zlg,Banerjee:2021cly,Banerjee:2020vnt,Banerjee:2021dlm}.   }
	  
	  Our results pertain to celestial operators representing boost eigenstates of massless particles, constructed from momentum eigenstates by a Mellin transformation with respect to energy. In particular, we do not investigate the boost eigenstates of massless particles that are constructed with an additional shadow or light transform,  for which considerations of locality become murky. We also do not treat boost eigenstates of massive particles,  whose local behavior is similarly less clear.  Our results include coefficients for OPEs involving massless fermions and scalars as well as those arising from non-minimal higher derivative gauge and gravitational couplings  such as $F^3$, $R^3$, etc.

	  Amazingly, Poincar\'e symmetry, the first few leading soft theorems employed in \cite{Pate:2019lpp}, and their local enhancements do not exhaust the list of known symmetries of asymptotically flat spacetimes.  In addition, the symmetries arising from the first few leading soft theorems are at the very least augmented  by an infinite tower  \cite{Guevara:2021abz}.  These arise from further subleading soft theorems \cite{Li:2018gnc,Hamada:2018vrw} and, with an appropriate labelling, generate  a ${\rm w}_{1+ \infty}$ Kac-Moody symmetry \cite{Strominger:2021lvk}.  More precisely, these operators form a current algebra for the global ``wedge subalgebra"\footnote{For example, ${\rm SL}(2, \mathbb{R})$ is the wedge subalgebra of (one copy of) Virasoro.} of ${\rm w}_{1+ \infty}$.

	  The latter part of the paper is dedicated to showing that the OPE coefficients derived from Poincar\'e symmetry respect the algebra generated by the infinite tower of conformally soft graviton currents in minimally-coupled\footnote{Corrections to the generalized soft theorems from higher-derivative effective field operators are absent in minimally-coupled theories of gravity \cite{Elvang:2016qvq,Laddha:2017vfh}.  It would be very interesting and nontrivial to understand the deformation of the symmetry algebra and ensuing OPE constraints that are induced by these corrections.  We leave this open problem to a future investigation.} tree-level gravitational theories.  To do so, we explicitly construct currents and show that they generate the action of ${\rm w}_{1 +\infty}$ on  massless celestial operators. The construction of the currents involves a light transform \cite{Kravchuk:2018htvh}, which has played an increasingly important role in celestial holography \cite{Atanasov:2021cje,Sharma:2021gcz, Strominger:2021lvk, Alfredonew}.  As before, our approach is manifestly 2D conformally covariant.  In this approach, the ${\rm SL}(2, \mathbb{R})$ subalgebra of ${\rm w}_{1 +\infty}$  is readily identified as the right-moving ${\rm SL}(2, \mathbb{R})$ global conformal symmetry upon analytic continuation to the celestial torus \cite{Atanasov:2021oyu}.  Another auxiliary result in this work is a simple and explicit construction of the scalar primary operator that generates supertranslations and was previously discussed in \cite{Barnich:2017ubf,Fotopoulos:2019vac,Fotopoulos:2020bqj,Pasterski:2021fjn}.

	  This paper is organized as follows.  In Section \ref{sec:poincare}, we briefly review the Poincar\'e  symmetry generators and their action on massless celestial primary operators.    Our starting point in Section \ref{sec:ope_sym} is the general ansatz for the leading behavior at tree-level in a holomorphic limit  of an operator product expansion between massless celestial primaries that was given in \cite{Pate:2019lpp}.  We then determine the action of 4D translation symmetry on global conformal primaries and descendants, and use this action to obtain constraints on the OPE coefficients of primaries and descendants. These symmetry constraints are found to take the form of recursion relations, which can be solved systematically. We present a closed-form solution, which is unique up to overall constants of proportionality that are directly related to three-point coupling constants. In Section \ref{sec:OPEcol}, we  use a BCFW shift to isolate the leading holomorphic collinear singularity of tree-level momentum-space amplitudes and subsequently transform the result to a boost eigenstate basis.  The  resulting OPE coefficients agree with those derived from symmetry in Section \ref{sec:ope_sym}.  In Section \ref{sec:w-gen}, we construct currents that generate the ${\rm w}_{1+ \infty}$ symmetry of \cite{Strominger:2021lvk}, compute their OPE with massless celestial primaries, and determine the symmetry action on massless celestial primaries. In Section \ref{sec:w-ope}, we show that the OPEs derived from Poincar\'e in Section \ref{sec:ope_sym} also respect the algebra generated by the infinite tower of currents constructed in Section \ref{sec:w-gen}.  Conventions are summarized in Appendix \ref{appA}.   In Appendix \ref{app:old-sym}, we derive the symmetry action on massless celestial primaries of the conformally soft graviton currents studied in \cite{Guevara:2021abz} and demonstrate that the OPE is also invariant under this presentation of the transformations. Appendix \ref{app:w-proof} contains a proof  that the currents constructed in Section \ref{sec:w-gen} generate the action of ${\rm w}_{1+\infty}$ when acting on massless celestial primary operators.  \\
	  
	  \noindent \textbf{Note added:} After the completion of this work, we learned of \cite{Jiang:2021ovh}, which contains some overlapping results.

   \section{Poincar\'e in Celestial Amplitudes}
    \label{sec:poincare}
    
    In this section, we briefly review Poincar\'e symmetry in the context of celestial amplitudes, studied for example in \cite{Stieberger:2018onx}.  The Lorentz subgroup of Poincar\'e in four dimensions is realized by the global conformal group in two dimensions. As familiar from 2D CFT, the generators $L_n$, ${\bar L}_n$, $n= 0, \pm 1$  respect  
    \begin{equation}
        \begin{split}
            \left[L_m, L_n\right] = (m-n)L_{m+n}, \quad \quad \quad 
                 \left[\bar{L}_m, \bar{L}_n\right] = (m-n)\bar{L}_{m+n},
        \end{split}
    \end{equation}
   and the symmetry action on  primary operators takes the form 
   \begin{equation}
       \begin{split}
           \left[L_m, \mathcal{O}_{h, \bh}(z, \bz)  \right]
            & = z^{m} \left((m+1)h +z  \partial_z\right)\mathcal{O}_{h, \bh}(z, \bz),
       \end{split}
   \end{equation}
   \begin{equation}\label{eq:sl2r-r}
       \begin{split}
           \left[\bar{L}_m, \mathcal{O}_{h, \bh}(z, \bz)  \right]
            & = \bz^{m} \left((m+1)\bh +\bz  \partial_\bz\right)\mathcal{O}_{h, \bh}(z, \bz).
       \end{split}
   \end{equation} 
    Here and henceforth, $\co_{h , \bh } (z , \bz )$ represents an outgoing particle of definite boost weight $\Delta = h+ \bh$ and helicity $s = h- \bh$.  Throughout we assume that operators of arbitrary (i.e.  non-integer) conformal weight are constructed by Mellin-transforming momentum-space massless particles without an additional shadow or light transform.  We also do not consider massive particles.

    The subgroup of 4D translations can also be represented by charges in the 2D theory, which act on celestial primary operators via the transformation \cite{Stieberger:2018onx} 
    \begin{equation}\label{eq:translations}
        \begin{split}
            \left[P_{m,n},\mathcal{O}_{h, \bh}(z, \bz) \right]
                =\frac{1}{2} z^{m+\frac{1}{2}}
                    \bz^{n+\frac{1}{2}}\mathcal{O}_{h+ \frac{1}{2}, \bh+ \frac{1}{2}}(z, \bz),
                    \quad\quad \quad 
                    m, n = \pm \frac{1}{2}.
        \end{split}
    \end{equation}
    The generators $P_{m,n}$ transform under global conformal symmetry like modes of a primary operator with weight $(h, \bh) = \left(\frac{3}{2}, \frac{3}{2}\right)$ \cite{Barnich:2017ubf,Fotopoulos:2019vac}
    \begin{equation} \label{lp-mode}
        \begin{split}
            \left[L_k, P_{m,n}\right] &= \left(\frac{1}{2}k-m\right)P_{m+k,n}, \quad \quad \quad 
             \left[\bar{L}_k, P_{m,n}\right] = \left(\frac{1}{2}k-n\right)P_{m,n+k},
        \end{split}
    \end{equation}
    and form a closed algebra with the global conformal generators.\footnote{The supercurrent in a 2D superconformal field theory is a familiar example of a weight $h =\frac{3}{2}$ primary operator that admits a mode decomposition in which the $\pm \frac{1}{2}$ modes form a closed subalgebra with the global conformal generators.}  In gravitational theories, the celestial primary operator weight of $(\frac{3}{2}, \frac{3}{2})$ can be identified with a particular asymptotic state of the graviton.  In Section \ref{sec:w-gen}, we explicitly construct this operator as the light transform of a $\Delta =1$ conformally soft graviton.\footnote{This primary was previously discussed in \cite{Barnich:2017ubf,Fotopoulos:2019vac,Fotopoulos:2020bqj,Pasterski:2021fjn} but  not directly constructed with a light transform.}

    In fact, it is also compatible with  Poincar\'e to regard the translation generators as the modes of a primary operator in which the left, right, or both conformal weights are $-\frac{1}{2}$ as opposed to $\frac{3}{2}$.  Such primary operators can be constructed by taking various light or shadow transforms of the $\Delta = 1$ conformally soft graviton. However, these modes do not automatically form a closed algebra with the global conformal generators and require an additional assumption of a truncated mode expansion as presented, for example, in \cite{Banerjee:2020zlg,Guevara:2021abz}.  More details on this alternative mode expansion are provided in  Appendix \ref{app:old-sym}.

\section{OPEs from Poincar\'e Symmetry}
    \label{sec:ope_sym}
     Throughout, we take $z$ and $\bar{z}$ to be independent and continue the Lorentz symmetry ${\rm SL}(2, \mathbb{C})$ to ${\rm SL}(2, \mathbb{R})_L \otimes {\rm SL}(2, \mathbb{R})_R$, following for example \cite{Pate:2019lpp, Guevara:2021abz}.  Working to leading order in a holomorphic limit $z \to 0$ with $\bz$ fixed,  the contribution from primary operators in the celestial OPE at tree-level has been demonstrated \cite{Pate:2019lpp} to take the form\footnote{It would be interesting to provide a general derivation of the leading holomorphic behavior from symmetry, as  given in \cite{Banerjee:2020vnt,Banerjee:2020zlg} for the MHV sector of gauge theory and gravity.}  
	\begin{equation} \label{OPEgen}
	    \begin{split}
	        \mathcal{O}_{h_1, \bh_1}(z, \bz)\mathcal{O}_{h_2,\bh_2}(0,0)
	             \sim \frac{1}{z}\sum_{p} C_{p} (  \bh_1,  \bh_2) 
	                    \bz^{p} 
	                    \mathcal{O}_{h_1+h_2-1, \bh_1+\bh_2+p}(0, 0).
	    \end{split}
	\end{equation}
	Here $p = d_V-4$, where $d_V$ is the bulk dimension of the three-point interaction coupling particles of spin (4D helicity) $s_1$, $s_2$ and $p+1-s_1-s_2$.   The OPE coefficients $C_p$ also depend on the left conformal weights $h_i$. However, our goal in this section will be to determine the entire dependence of the OPE coefficients on the conformal dimensions $\Delta_i$, but not the dependence on spin $s_i$. Without loss of generality, we can instead regard the OPE coefficients as functions of spin $s_i$ and right conformal weights $\bh_i$ and determine the dependence on $\bh_i$.  We leave the dependence on spin implicit in our notation.

	Our method in this section for determining the OPE coefficient $C_{p} ( \bh_1,\bh_2)$ will be to impose constraints arising from Poincar\'e symmetry reviewed in the previous section.  The symmetry action \eqref{eq:translations} non-trivially mixes ${\rm SL}(2, \mathbb{R})_R$ primaries and descendants, so when deriving constraints on the OPE, we need to include the contribution from right-moving descendants.\footnote{We find that  it is sufficient to study the action of $P_{-\frac{1}{2}, \pm\frac{1}{2}}$.  These charges do not mix ${\rm SL}(2, \mathbb{R})_L$ primaries and descendants, so we can consistently analyze the leading term in a holomorphic limit  at tree level on its own.}  However, the symmetry transformations \eqref{eq:translations}  preserve the spin of the primary, so we need only focus on the contribution of fixed $p$ (equivalently of a coupling of fixed dimension $d_V$).   We therefore begin with the ansatz
	\begin{equation} \label{ansatz}
	    \begin{split}
	        \mathcal{O}_{h_1, \bh_1}(z,\bz)\mathcal{O}_{h_2,\bh_2}(0,0)
	             \sim \frac{1}{z}\sum_{m = 0}^\infty C^{(m)}_{p} ( \bh_1,  \bh_2) 
	                    \bz^{p+m} 
	               \bar{\partial}^m    \mathcal{O}_{h_1+h_2-1, \bh_1+\bh_2+p}(0,0),
	    \end{split}
	\end{equation}
	where $C^{(m)}_{p} ( \bh_1,  \bh_2) $ denotes the OPE coefficient of the $m$th right-moving descendant. As written, the OPE \eqref{ansatz} already respects the symmetries generated by $\bar{L}_{-1}$ and $\bar{L}_0$. 
	
    As a warm-up, we now review how the symmetry generated by $\bar{L}_1$ constrains the  OPE coefficients of descendants $C^{(m>0)}_{p} (\bh_1, \bh_2)$ in terms of the OPE coefficient of the primary  $C^{(0)}_{p} (\bh_1,\bh_2)$.    First, recall that the action of $\bar{L}_1$ on a primary operator  is given by
    \begin{equation}
         \left[\bar{L}_1 , \mathcal{O}_{h,\bh}(z, \bz) \right]
        = \left(2 \bh \bz  + \bz^2 \partial_\bz\right)\mathcal{O}_{h,\bh}(z, \bz).
    \end{equation}
    The action on descendants is then given by\footnote{This could have equivalently been derived from the ${\rm SL}(2, \mathbb{R})_R$ algebra \eqref{eq:sl2r-r}.
   In particular, it is helpful to note that
   $
    \bar{L}_1 \bar{L}_{-1}^m = \bar{L}_{-1}^m \bar{L}_1 + 2 m  \bar{L}_{-1}^{m-1} \bar{L}_0 +  m (m-1) \bar{L}_{-1}^{m-1} .
   $}
     \begin{equation}
        \begin{split}
         \left[\bar{L}_1 , \partial_{\bz}^m\mathcal{O}_{h,\bh}(z, \bz) \right]
                & =\partial_{\bz}^m \left[\bar{L}_1 , \mathcal{O}_{h,\bh}(z, \bz) \right]\\
                 & =\partial_{\bz}^m  
                \left(2 \bh \bz  + \bz^2 \partial_\bz\right)\mathcal{O}_{h,\bh}(z, \bz) \\
                 &= \left( \left(2 \bh \bz + \bz^2 \partial_\bz\right)\partial_\bz^{m} 
                + 2m \left(\bh + \bz\partial_\bz \right)\partial_\bz^{m-1} 
                + m (m-1) \partial_\bz^{m-1}\right)\mathcal{O}_{h,\bh}(z, \bz).
         \end{split}
    \end{equation}
    Note  that when $\bz=0$, the action dramatically simplifies to
     \begin{equation}
        \begin{split}
            \left[\bar{L}_1 ,
                \bar{\partial}^m \mathcal{O}_{h,\bh}(0,0)\right]
                 &=  m \left(2 \bh+m -1\right)  \bar{\partial}^{m-1}\mathcal{O}_{h,\bh}(0,0) .
        \end{split}
    \end{equation}
    
    To constrain the OPE, we first determine the action of $\bar{L}_1$ on the left-hand side of \eqref{ansatz}.  Noting that $\bar{L}_1$ annihilates a primary at the origin, we find 
    \begin{equation}
        \begin{split}
           \left[ \bar{L}_1  ,
             \mathcal{O}_{h_1, \bh_1}(z, \bz)\mathcal{O}_{h_2,\bh_2}(0,0)\right]
            & = \left(2 \bh_1 \bz+ \bz^2 \partial_\bz\right)
            \mathcal{O}_{h_1, \bh_1}(z, \bz)\mathcal{O}_{h_2,\bh_2}(0,0)
            \\
       & \sim 
        \frac{\bz^{p+1}}{z}\sum_{m = 0}^\infty 
	                    \left( 2 \bh_1 +p+m\right) 
        C^{(m)}_{p} (  \bh_1, \bh_2)
	                    \bz^{m}
	                \bar{\partial}^m    \mathcal{O}_{h_1+h_2-1, \bh_1+\bh_2+p}(0,0).
        \end{split}
    \end{equation}
    Equating this with the action of $\bar{L}_1$ on the right-hand side of \eqref{ansatz}, 
    \begin{equation}
        \begin{split}
            \frac{1}{z} &\sum_{m = 0}^\infty C_p^{(m)} (  \bh_1, \bh_2)
							 	\bz^{m+p} \left[\bar{L}_1, \bar{\partial}^m \co_{h_1+h_2-1, \bh_1+\bh_2+p}(0,0)\right]\\
								 &  =\frac{\bz^{p+1}}{z} \sum_{m = 0}^\infty
								   (m{+}1) \left(2  \bh_1{+}2\bh_2{+}2p{+}m \right) 
								    C_p^{(m+1)} (  \bh_1, \bh_2)
							 	\bz^{m}\bar{\p}^{m}\co_{h_1+h_2-1, \bh_1+\bh_2+p}(0,0), 
        \end{split}
    \end{equation}
     we obtain the constraint
     \begin{equation} \label{L1-const}
          \left( 2 \bh_1 +p+m\right)
            C^{(m)}_{p} (  \bh_1, \bh_2)
             =  (m+1) \left(2  \bh_1+2\bh_2+2p+m \right)   C_p^{(m+1)} (  \bh_1, \bh_2).
     \end{equation}
     As expected, the symmetry constraint associated to $\bar{L}_1$ is a recursion relation in $m$, relating the OPE coefficients of descendants to that of the primary. 
     
     Next, we apply this analysis to the translation generators.   $P_{-\frac{1}{2}, -\frac{1}{2}}$ was used in \cite{Pate:2019lpp} to show that the coefficients of the primary must obey  
     \begin{equation} \label{p-1/2-1/2}
         C_{p}^{(0)}(\bh_1+\tfrac{1}{2}, \bh_2)
            +C_{p}^{(0)}(\bh_1, \bh_2+\tfrac{1}{2})
                 = C_{p}^{(0)}(\bh_1, \bh_2).
     \end{equation}
    This constraint alone is not sufficient to determine the OPE coefficient uniquely. However, when combined with the constraint from $P_{-\frac{1}{2}, \frac{1}{2}}$, the $\bh$ dependence is  fixed uniquely, as we will now show. 
     
    To begin,  we recall the action of $P_{-\frac{1}{2}, \frac{1}{2}}$ on a primary operator: 
     \begin{equation}
         \left[P_{-\frac{1}{2}, \frac{1}{2}},  \co_{h , \bh } (z , \bz )\right] 
							= \frac{1}{2} \bz			 	\co_{h +\frac{1}{2},  \bh +\frac{1}{2}} (z , \bz ).
     \end{equation}
    The action of $P_{-\frac{1}{2}, \frac{1}{2}}$ on a descendant takes the form
	\begin{equation}
	   \begin{split}
					\left[ P_{-\frac{1}{2}, \frac{1}{2}} , \partial_\bz^m \co_{h, \bh}(z, \bz)\right]
					    & = \partial_\bz^m\left[ P_{-\frac{1}{2}, \frac{1}{2}},  \co_{h, \bh}(z,
					    \bz)\right]
					     = \frac{1}{2}\partial_\bz^m \left(\bz   \co_{h+\frac{1}{2}, \bh+ \frac{1}{2}}(z, \bz )\right).
				\end{split}
	\end{equation} 
	This action can be equivalently derived from the mode algebra \eqref{lp-mode}. 	When $\bz = 0$, the action  	simplifies to
	\begin{equation} \label{s32-simp}
	    \begin{split}
	        \left[ P_{-\frac{1}{2}, \frac{1}{2}},  \bar{\partial}^m \co_{h, \bh}(0,0)\right]
						& =   \frac{1}{2}m\bar{\partial}^{m-1} \co_{h+\frac{1}{2}, \bh+ \frac{1}{2}}(0,0).
	    \end{split}
	\end{equation}	
	 Note that $P_{-\frac{1}{2}, \frac{1}{2}}$, like $\bar{L}_1$, maps a level-one descendant to the primary. 
	
    Next, acting on the left-hand side of \eqref{ansatz} with $P_{-\frac{1}{2}, \frac{1}{2}}$ and using the fact that it annihilates an operator at the origin,  we find
    \begin{equation} \label{lhs-sb}
        \begin{split}
           \left[ P_{-\frac{1}{2}, \frac{1}{2}},
             \mathcal{O}_{h_1, \bh_1}(z, \bz)\mathcal{O}_{h_2,\bh_2}(0,0)  \right] 
              & =\frac{1}{2} \bz
            \mathcal{O}_{h_1+\frac{1}{2}, \bh_1+\frac{1}{2}}(z, \bz)\mathcal{O}_{h_2,\bh_2}(0,0)
            \\
        & 
        \sim 
       \frac{1}{2} \frac{\bz^{p+1}}{z}\sum_{m = 0}^\infty  
        C^{(m)}_{p} (  \bh_1+ \tfrac{1}{2}, \bh_2)
	                    \bz^{m}
	                \bar{\partial}^m    \mathcal{O}_{h_1+h_2-\frac{1}{2}, \bh_1+\bh_2+p+\frac{1}{2}}(0,0).
        \end{split}
    \end{equation}
    To determine the action of $P_{-\frac{1}{2}, \frac{1}{2}}$ on
    the right-hand side of     \eqref{ansatz}, we use \eqref{s32-simp} and find
    \begin{equation}\label{rhs-sb}
        \begin{split}
            \frac{1}{z}\sum_{m = 0}^\infty C^{(m)}_{p} ( \bh_1,  \bh_2) 
	                  &  \bz^{p+m} 
	                \left[P_{-\frac{1}{2}, \frac{1}{2}} , \bar{\partial}^m    \mathcal{O}_{h_1+h_2-1, \bh_1+\bh_2+p}(0,0)\right]\\
	    &=  \frac{1}{2} \frac{\bz^{p+1}}{z}\sum_{m = 0}^\infty
	            (m+1)
	            C^{(m+1)}_{p} ( \bh_1,  \bh_2) 
	            \bz^{m}  \bar{\partial}^m    \mathcal{O}_{h_1+h_2-\frac{1}{2}, \bh_1+\bh_2+p+\frac{1}{2}}(0,0) .
        \end{split}
    \end{equation}
    Equating \eqref{lhs-sb} and \eqref{rhs-sb}, we obtain the constraint
    \begin{equation} \label{s32-only}
        \begin{split}
        C^{(m)}_{p} (  \bh_1+ \tfrac{1}{2}, \bh_2) = (m+1)C^{(m+1)}_{p} ( \bh_1,  \bh_2) ,
	   \end{split}
    \end{equation}
    which involves a recursion relation in both $m$ and $\bh_1$.   By combining \eqref{s32-only} with the constraint \eqref{L1-const} from $\bar{L}_1$, we obtain a constraint for fixed $m$: 
     \begin{equation}
        \begin{split}  
	       \left( 2 \bh_1 +p+m\right)
            C^{(m)}_{p} (  \bh_1, \bh_2)
             =    \left(2  \bh_1+2\bh_2+2p+m \right)   C^{(m)}_{p} (  \bh_1+ \tfrac{1}{2}, \bh_2).
	   \end{split}
    \end{equation}  
    Setting $m = 0$, we find the OPE coefficients of the primaries must respect the recursion relation
    \begin{equation} \label{rec-1}
         \left( 2 \bh_1 +p\right)
            C^{(0)}_{p} (  \bh_1, \bh_2)
             =    \left(2  \bh_1+2\bh_2+2p \right)   C^{(0)}_{p} (  \bh_1+ \tfrac{1}{2}, \bh_2).
    \end{equation}
    Combining this with the constraint \eqref{p-1/2-1/2} from $P_{-\frac{1}{2}, -\frac{1}{2}}$, we obtain a constraint that is similar to \eqref{rec-1}, but involving recursion in $\bar{h}_2$ instead of $\bar{h}_1$:
    \begin{equation} \label{rec-2}
         \left( 2 \bh_2 +p\right)
            C^{(0)}_{p} (  \bh_1, \bh_2)
             =    \left(2  \bh_1+2\bh_2+2p \right)   C^{(0)}_{p} (  \bh_1, \bh_2+ \tfrac{1}{2}).
    \end{equation}
    As  shown  in \cite{Pate:2019lpp}, the two relations \eqref{rec-1} and \eqref{rec-2}, together with suitable assumptions about boundedness and analyticity in $\bh_1$ and $\bh_2$,  uniquely fix the primary OPE coefficient to take the form
    \begin{equation}
        C^{(0)}_p (\bh_1, \bh_2) 
            = \gamma_{p}^{s_1, s_2} B(2 \bh_1+p, 2 \bh_2+p).
    \end{equation} Here $\gamma_p^{s_1,s_2}$ is an overall normalization that is related to the coupling coefficient of the dimension $d_V = p+4$ three-point interaction between particles of spin $s_1$, $s_2$, and $p+1-s_1-s_2$.  $B(x,y)$ is the Euler beta function, which can be expressed as the following ratio 
	 of gamma functions:
	\begin{equation}
	    B(x,y) = \frac{\Gamma (x) \Gamma (y)}{\Gamma (x+y)}.
	\end{equation}
	
	Finally, the OPE coefficients of the descendants can be determined from the OPE coefficient of the primary  using the $\bar{L}_1$ recursion relation in $m$ \eqref{L1-const}. One could equivalently use the $P_{-\frac{1}{2}, \frac{1}{2}}$ recursion \eqref{s32-only}. These coefficients also admit a simple, closed-form  expression:  
	\begin{equation}
        \label{mainresult2}
	    C^{(m)}_p (\bh_1, \bh_2) 
            = \gamma_{p}^{s_1, s_2}\frac{1}{m!} B(2 \bh_1+p+m, 2 \bh_2+p).
	\end{equation}
    This is a central result of the paper.  As we explicitly verify from collinear limits in the following section, our formula \eqref{mainresult2} is valid for any three-point interaction $\gamma_p^{s_1,s_2}$ of bulk scaling dimension $d_V = p+4$. In particular, it holds for three-point interactions involving fermions as well as those arising from
    non-minimal coupling of gravitons and gluons (such as for example $\phi F^2$, $F^3$, $RF^2$ and $R^3$).   
    
    The two main ingredients in the derivation of this result are Poincar\'e symmetry and the ansatz \eqref{OPEgen}.  Poincar\'e is assumed to be an exact symmetry of the scattering problem, uncorrected by loops or effective field theory operators. On the other hand, the ansatz \eqref{OPEgen} relies on a tree-level argument, which can receive corrections.  For example, in gauge theories with massless matter, the ansatz should exhibit a branch cut extending from $z = 0$ to reproduce the loop corrections in momentum space that are known to appear at every order in perturbation theory \cite{Feige:2014wja,Kosower:1999xi}.  In particular, the singularity structure at one-loop includes terms of the form $\sim \frac{1}{z} \log z$.

\section{OPEs from Collinear Limits}
		\label{sec:OPEcol}

	In this section we verify that the OPE coefficients derived from symmetry in the previous section match those from collinear limits.  To obtain OPE coefficients from collinear limits in momentum space, we first derive the leading holomorphic collinear singularity from a BCFW shift of the momentum space amplitude.  Similar analyses were performed in \cite{Guevara:2021abz,Guevara:2019ypd} to establish other properties of celestial amplitudes.   
		
	It will be convenient to introduce spinor helicity variables, parametrized by  an energy $\omega$ and a point $z$ on the complex plane:
		\be
			\begin{split}
				\lambda  =\eps   \sqrt{\omega} \left(\begin{matrix} 1 \\z\end{matrix}\right), \quad \quad \quad
					\tilde \lambda = \sqrt{\omega} \left(\begin{matrix} 1 \\ \bz\end{matrix}\right).
			\end{split}
		\ee
		Here $\eps = \pm1$ for outgoing/incoming particles. An overview of our conventions can be found in Appendix \ref{appA}. These variables obey 
		\be
			\langle ij \rangle = - \eps_i \eps_j  \sqrt{\omega_i \omega_j} z_{ij}, \quad \quad \quad 
				[ij] =\sqrt{\omega_i \omega_j} \bz_{ij}.
		\ee
		Our goal will be to determine the leading $z_{12}$ pole in a tree-level $n$-particle scattering amplitude. 
		
		To determine this, suppose particles $1$, $2$, and $k \neq 1,2$ are all outgoing and consider the BCFW deformation \cite{Britto:2005fq}
		\be
			\hat \lambda_1  = \lambda_1  + z \lambda_k, \quad \quad \quad 
				\hat{\tilde{\lambda}}_k =\tilde \lambda_k - z \tilde \lambda_1.
		\ee
		As usual, the undeformed amplitude is extracted from the deformed one by contour integration: 
		\be
			{\bf A}_n ( \lambda_i, \tilde \lambda_i)
				 = \oint_{|z|< \eps} \frac{dz}{2 \pi i } \frac{1}{z} {\bf A}_n ( \hat \lambda_i, \hat {\tilde \lambda}_i).
		\ee
		Deforming the $z$ contour to encircle poles in ${\bf A}_n ( \hat \lambda_i, \hat {\tilde \lambda}_i)$, we note that simple $z_{12}$-poles in the undeformed amplitude ${\bf A}_n ( \lambda_i, \tilde \lambda_i)$ arise from simple $z$-poles in the deformed amplitude ${\bf A}_n ( \hat \lambda_i, \hat {\tilde \lambda}_i)$ at $z = \xi z_{12}$: 	\be \label{BCFW1}
			\begin{split}
				{\bf A}_n ( \lambda_i, \tilde \lambda_i)
				 & = - \frac{1}{\xi z_{12}}{\rm Res}_{z = \xi z_{12}} \left[{\bf A}_n ( \hat \lambda_i, \hat {\tilde \lambda}_i)\right]
				 	+ \co (z_{12}^0),
			\end{split}
		\ee
		 where $\xi$ is a to-be-determined parameter.  Only factors of the form $\frac{1}{\langle \hat 1 \hat 2\rangle}$ can give rise to poles at $z =  \xi z_{12}$.   From 
		 \begin{equation}
		     \langle \hat 1 \hat 2\rangle
		         =  \langle 12 \rangle + z \langle k2 \rangle 
		         =\sqrt{\omega_k \omega_2} z_{2k}
		            \left(z- \sqrt{\frac{\omega_1}{\omega_k}}\frac{z_{12}}{z_{2k}}\right),
		 \end{equation} 
		we identify
		\be
			\xi = \sqrt{\frac{\omega_1}{\omega_k}}\frac{1}{z_{2k}}.
		\ee
		It will be useful to note that
		\be
			(\hat p_1 + \hat p_2)^2 =\langle \hat 1 \hat 2\rangle [\hat 1 \hat 2]
				= \frac{2}{\xi} \omega_1 \omega_2 \bz_{12} \left(z- \xi z_{12} \right).
		\ee
		On the pole $z = \xi z_{12}$, the amplitude factorizes so that the residue is given by a sum over products of lower point amplitudes:
		\be \label{res4323}
			\begin{split}
				{\rm Res}_{z = \xi z_{12}}  \left[{\bf A}_n ( \hat \lambda_i, \hat {\tilde \lambda}_i)\right]
							& =\sum_{s_I} {\rm Res}_{z = \xi z_{12}} \left[\frac{{\bf A}^{-s_I}_3(\hat p_1, \hat p_2, -\hat p_1-\hat p_2) {\bf A}^{s_I}_{n-1}(\hat p_1+\hat p_2, \cdots )}{(\hat p_1+ \hat p_2)^2}\right]\\
						& =\frac{\xi}{2 \omega_1 \omega_2 \bz_{12}} \sum_{s_I} 
							{\bf A}^{-s_I}_3\left( \sqrt{\frac{\omega_1}{ \omega_2}} \frac{z_{1k}}{z_{2k}} \lambda_2, \tilde \lambda_1;
								\lambda_2, \tilde \lambda_2;
								-\sqrt{1+\frac{ \omega_1  }{\omega_2}}\lambda_2,
								\frac{\sqrt{  \omega_1 } \frac{z_{1k}}{z_{2k}}  \tilde \lambda_1 +\sqrt{ \omega_2 }  \tilde \lambda_2}{\sqrt{\omega_1+\omega_2}}	
									\right)\\& \quad \quad \quad \quad \quad \quad \quad \quad \quad \times
									 {\bf A}^{s_I}_{n-1} \left(\sqrt{1+\frac{ \omega_1  }{\omega_2}}\lambda_2,
								\frac{\sqrt{  \omega_1 } \frac{z_{1k}}{z_{2k}}  \tilde \lambda_1 +\sqrt{ \omega_2 }  \tilde \lambda_2}{\sqrt{\omega_1+\omega_2}}	; \cdots \right).
			\end{split}
		\ee
		Here $-s_I$ labels the helicity of the third particle appearing in the three-point amplitude ${\bf A}_3$ with particle 1 and 2.  	The sign is chosen such that it labels the helicity of the particle in $(n-1)$-point amplitude.  All other	labels are suppressed but implicitly included in the sum. 
		
		To simplify further, we need the explicit form of the three-point amplitude: 
		\be
			\begin{split}
			{\bf A}_3 (\lambda_1, \tilde \lambda_1;
				\lambda_2, \tilde \lambda_2;\lambda_3, \tilde \lambda_3)
				& = \left \{ \begin{matrix}
						 g_{123} [12]^{s_1+s_2-s_3}[32]^{s_2+s_3-s_1}[13]^{s_3+s_1-s_2}, & s_1+s_2+s_3>0\\
						 \tilde g_{123} 
							\langle 12 \rangle^{s_3-s_1-s_2}
								\langle 32\rangle^{s_1-s_2-s_3}\langle 13 \rangle^{s_2-s_3-s_1}, & s_1+s_2+s_3<0
					\end{matrix}
					\right. .
			\end{split} 
		\ee
		Since all $\lambda_i$ appearing in the three-point amplitudes in \eqref{res4323} are proportional to $\lambda_2$, only three-point 
		amplitudes with $s_1+s_2+s_3>0$ will contribute and \eqref{res4323} becomes  
		\be  
			\begin{split}
				{\rm Res}_{z = \xi z_{12}}  \left[{\bf A}_n ( \hat \lambda_i, \hat {\tilde \lambda}_i)\right] 
						& =\frac{\xi}{2 } \sum_{s_I} 
							 g_{12I} ~  \omega_1^{s_2-s_I-1 } \omega_2^{s_1 -s_I -1} (\omega_1+\omega_2 )^{s_I}
							  \bz_{12}^{s_1+s_2-s_I-1}  
							 \left(\frac{z_{1k}}{z_{2k}}\right)^{s_2-s_I-s_1} 
									\\& \quad \quad \quad \quad \quad \quad \quad \quad \quad \times
									 {\bf A}^{s_I}_{n-1} \left(\sqrt{1+\frac{ \omega_1  }{\omega_2}}\lambda_2,
								\frac{\sqrt{  \omega_1 } \frac{z_{1k}}{z_{2k}}  \tilde \lambda_1 +\sqrt{ \omega_2 }  \tilde \lambda_2}{\sqrt{\omega_1+\omega_2}}	; \cdots \right).
			\end{split}
		\ee
		Substituting this result in \eqref{BCFW1} and expanding to leading order in $z_{12}$, we find
		\be  
			\begin{split}
				{\bf A}_n (\omega_i, z_i, \bz_i)
				 & = - \frac{1}{ 2 z_{12}} \sum_{s_I} 
							 g_{12I}   \omega_1^{s_2-s_I-1 } \omega_2^{s_1 -s_I -1} (\omega_1+\omega_2 )^{s_I}
							  \bz_{12}^{s_1+s_2-s_I-1}    
									\\& \quad \quad \quad\quad \quad \quad \quad \quad \quad \quad \quad \times
									 {\bf A}^{s_I}_{n-1} \left(\omega_1+\omega_2, z_2, \frac{\omega_1 \bz_1+\omega_2\bz_2}{\omega_1+\omega_2};  \cdots \right)
				 	+ \co (z_{12}^0).
			\end{split}
		\ee
		
	The associated relation among celestial amplitudes $\mathcal{A}$ then immediately follows from their 
	relation to momentum space amplitudes ${\bf A}$:
	\begin{equation}
	    \mathcal{A}_n (\Delta_i, z_i, \bz_i)
	        \equiv 
	        \Big ( \prod_{j = 1}^n \int_0^\infty \frac{d \omega_j}{ \omega_j} \omega_j^{\Delta_j} 
	        \Big) {\bf A}_n(\omega_i, z_i, \bz_i).
	\end{equation}
	Explicitly, we find
	\begin{equation}
	    \begin{split}
	         \mathcal{A}_n (\Delta_i, z_i, \bz_i)
	           & =- \frac{1}{ 2 z_{12}}  \Big ( \prod_{j = 3}^n \int_0^\infty \frac{d \omega_j}{ \omega_j} \omega_j^{\Delta_j} 
	        \Big) \sum_{s_I} 
							 g_{12I} \bz_{12}^{s_1+s_2-s_I-1} \\& 
							 \quad \quad \quad
	       \times \int_0^\infty \frac{d \omega_1}{ \omega_1} \omega_1^{\Delta_1}
	            \int_0^\infty \frac{d \omega_2}{ \omega_2} \omega_2^{\Delta_2}
	             \omega_1^{s_2-s_I-1 } \omega_2^{s_1 -s_I -1} (\omega_1+\omega_2 )^{s_I} 	\\& \quad \quad \quad\quad \quad \quad \quad \quad \quad \quad \quad \times
									 {\bf A}^{s_I}_{n-1} \left(\omega_1+\omega_2, z_2, \frac{\omega_1 \bz_1+\omega_2\bz_2}{\omega_1+\omega_2};  \cdots \right)
				 	+ \co (z_{12}^0).
	    \end{split}
	\end{equation}
	To obtain a factorized result in the boost weight basis, we make the change-of-variables
	\begin{equation}
	    \omega_1 = \omega t, \quad \quad \quad \omega_2= \omega (1-t),
	\end{equation}
	and find
	\begin{equation}
	    \begin{split}
	         \mathcal{A}_n (\Delta_i, z_i, \bz_i)
	           & =- \frac{1}{ 2 z_{12}} \sum_{s_I} 
							 g_{12I} \bz_{12}^{s_1+s_2-s_I-1}
	            \int_0^1 \frac{dt}{t (1-t)}   t^{\Delta_1 +s_2-s_I-1 } (1-t)^{\Delta_2+s_1 -s_I -1}  \\& 
							 \quad \quad  
	         \times \underbrace{\Big ( \prod_{j = 3}^n \int_0^\infty \frac{d \omega_j}{ \omega_j} \omega_j^{\Delta_j} 
	        \Big) \int_0^\infty \frac{d \omega}{ \omega} \omega^{\Delta_1+\Delta_2 +s_1+s_2-s_I-2}     {\bf A}^{s_I}_{n-1} \left(\omega
									 , z_2,\bz_2+ t\bz_{12};  \cdots \right)}
									 _{= \mathcal{A}_{n-1}
							  \left(\Delta_1+\Delta_2 +s_1+s_2-s_I-2, z_2,  \bz_2+ t\bz_{12};  \cdots \right)}
				 	+ \co (z_{12}^0).
	    \end{split}
	\end{equation}
	To perform the remaining $t$ integral, we Taylor-expand the result in $\bz_{12}$ and find
	\begin{equation}
	    \begin{split}
	         \mathcal{A}_n (\Delta_i, z_i, \bz_i)
	           & =- \frac{1}{ 2 z_{12}} \sum_{s_I} 
							 g_{12I} \bz_{12}^{s_1+s_2-s_I-1}
	            \int_0^1 \frac{dt}{t (1-t)}   t^{\Delta_1 +s_2-s_I-1 } (1-t)^{\Delta_2+s_1 -s_I -1}  \\& 
							 \quad \quad  
	         \times \sum_{m = 0}^\infty \frac{1}{m!} \bz_{12}^m t^m 
	         \partial_{\bz_{2}}^m
	         \mathcal{A}_{n-1}
							  \left(\Delta_1+\Delta_2 +s_1+s_2-s_I-2, z_2,  \bz_2 ;  \cdots \right) 
				 	+ \co (z_{12}^0)\\
			& =- \frac{1}{ 2 z_{12}} \sum_{s_I} 
			    \sum_{m = 0}^\infty 
							 g_{12I} 
	           \frac{1}{m!}  B\left(\Delta_1 +s_2-s_I-1+m ,\Delta_2+s_1 -s_I -1\right) \\& 
							 \quad \quad  
	         \times 
	         \bz_{12}^{s_1+s_2-s_I-1+m}\partial_{\bz_{2}}^m
	         \mathcal{A}_{n-1}
							  \left(\Delta_1+\Delta_2 +s_1+s_2-s_I-2, z_2,  \bz_2 ;  \cdots \right) 
				 	+ \co (z_{12}^0).	
	    \end{split}
	\end{equation}
	In the final line, we used the integral representation of the Euler beta function
	\begin{equation}
	    B(a, b) = \int_0^1 \frac{dt}{t(1-t)} t^a (1-t)^b.
	\end{equation}
	Thus, we have identified the OPE coefficient of the $m$th level descendant as
	\begin{equation}
	    C^{(m)}_p (\Delta_1, s_1; \Delta_2, s_2)
	         =
	         - \frac{1}{ 2  }  
							 g_{12I} 
	           \frac{1}{m!}  B\left(\Delta_1 +s_2-s_I-1+m ,\Delta_2+s_1 -s_I -1\right),
	\end{equation}
	which upon identifying $s_I = s_1+s_2 -p-1$ 
	 becomes  
	\begin{equation}
	    C^{(m)}_p (\Delta_1, s_1; \Delta_2, s_2)
	         =
	         - \frac{1}{ 2  }  
							 g_{12I} 
	           \frac{1}{m!}  B\left(\Delta_1 - s_1 + p+m ,\Delta_2-s_2 + p\right),
	\end{equation} 
     precisely matching the result \eqref{mainresult2} from the previous section. 
	Moreover, we find the overall proportionality constant $\gamma_p^{s_1, s_2}$
	is related to the three-point coupling by 
	\begin{equation} \label{eq:3ptcoupling}
	    \gamma_p^{s_1, s_2} =  - \frac{1}{ 2  }  
							 g_{12I} .
	\end{equation}

 \section{${\rm w}_{1+\infty}$ Currents}
    \label{sec:w-gen}
    
    In this section, we construct 2D currents that generate a ${\rm w}_{1+\infty}$ symmetry  action on celestial amplitudes \cite{Strominger:2021lvk}.  We introduce the light transform of a celestial operator
    \begin{equation} \label{def-lt}
        \begin{split}
            {\bf L}[\mathcal{O}_{h, \bh}] (z, \bz)
                \equiv
                        \int_{\mathbb{R}} \frac{d \bw}{2 \pi i}
                            \frac{1}{(\bz-\bw)^{2-2\bh}} \mathcal{O}_{h, \bh} (z, \bw).
        \end{split}
    \end{equation}     
   Positive helicity graviton currents are constructed by taking the following limit
    \begin{equation} \label{wcurrent-def}
        \begin{split}
           {\rm w}^q(z, \bz)
                 \equiv   \frac{1}{\kappa} (-1)^{2q}\Gamma (2q) \lim_{\varepsilon\to 0}
                    {\bf L}[\mathcal{O}_{3-q,1-q+\varepsilon }] (z, \bz), \quad \quad  
                        q = 1, \frac{3}{2}, 2, \frac{5}{2}, \cdots,
        \end{split}
    \end{equation}
    where $\kappa= \sqrt{32 \pi G}$ is the gravitational coupling constant.
       Note that $s = h-\bh = 2-\varepsilon$, so these currents constitute positive-helicity gravitons obtained from a limit in which spin (4D helicity) is continued from non-integer values. ${\bf L}[\mathcal{O}_{h, \bh}] (z, \bz)$ transforms under ${\rm SL}(2, \mathbb{R})_L \otimes {\rm SL}(2, \mathbb{R})_R$ like a  primary of weight $(h,1- \bh)$, which implies that   ${\rm w}^q$ transforms with ${\rm SL}(2, \mathbb{R})_L \otimes {\rm SL}(2, \mathbb{R})_R$  weight\footnote{Note that $q$ is a natural label for the anti-holomorphic sector because it is positive and parametrizes the weight of primaries under ${\rm SL}(2, \mathbb{R})_R$. On the other hand, the  ${\rm SL}(2, \mathbb{R})_L$ weight $h = 3-q$ is negative for sufficiently large $q$.  There may exist a more appropriate labelling such as the weight under the ${\rm SL}(2, \mathbb{R})$ generated by the ${\rm w}^q$ current Sugawara stress tensor. We leave this question to a future investigation.} 
    \begin{equation} \label{w-weight} 
        (h, \bh) = \left(3-q,q\right).
    \end{equation}

    To determine the action of the symmetry on massless celestial primaries, we first derive their OPEs with the  ${\rm w}^q$ operators. We begin with \eqref{ansatz} evaluated  with the OPE coefficients \eqref{mainresult2} and specialized to the  minimal coupling of a positive-helicity graviton to massless matter.  Importantly, we can phrase the minimal-coupling condition entirely in 2D language: for an OPE $\mathcal{O}_1 \mathcal{O}_2 \sim \mathcal{O}_3$ in which $\mathcal{O}_1$ is a positive-helicity graviton, the spin of $\mathcal{O}_2$ and $\mathcal{O}_3$ is required to be the same.  Noting that $s_2 = s_3$ implies $p = s_1 -1$, we find   
    \begin{equation}    \label{minimal-ope}
	    \begin{split}
	       \mathcal{O}_{h_1,\bh_1}&(z,\bz)\mathcal{O}_{h_2,\bh_2}(0,0)\\ 
	            & \sim \frac{\gamma^{s_1, s_2}_{s_1-1}}{z} \sum_{m = 0}^\infty\frac{1}{m!} B(2\bh_1+s_1 -1+m,     2\bh_2+s_1 -1)   \bz^{s_1-1+m} 
	               \bar{\partial}^m     \mathcal{O}_{h_2+ \bh_1+s_1-1, \bh_2+\bh_1+s_1-1}(0,0).
	    \end{split}
	\end{equation} 
    Eventually,  $\mathcal{O}_{h_1,\bh_1}(z,\bz)$ will be taken to be a positive helicity graviton.  For now, we leave the $\bar{h}_1$ and $s_1$ dependence arbitrary so that we can later take the limit  \eqref{wcurrent-def}. In fact,  \eqref{minimal-ope} also describes the minimal coupling of a positive helicity gauge boson to matter when $s_1 = 1$. Note  that nowhere in the derivation of \eqref{minimal-ope} from Poincar\'e did we assume that $p \in \mathbb{Z}$.

    Taking the light transform as defined in \eqref{def-lt}, this becomes    
	 \begin{equation}    \label{lt-ope}
	    \begin{split}
	        {\bf L}[ \mathcal{O}_{h_1,\bh_1}] (z,\bz)\mathcal{O}_{h_2,\bh_2}(0,0) 
	              &\sim -\frac{\gamma^{s_1,s_2}_{s_1-1}}{z} \sum_{m = 0}^\infty    
	              \frac{B(2\bh_1+s_1-1+m,     2\bh_2+s_1-1)\Gamma (2-2\bh_1-s_1-m)}{m!\Gamma(1-s_1-m)\Gamma (2-2\bh_1)}\\& \quad \quad \quad \quad \quad \quad 
	              \quad \quad \quad \quad \quad  \times 
	             \bz^{2\bh_1+s_1+m-2} 
	               \bar{\partial}^m     \mathcal{O}_{h_2+
	               \bh_1+s_1-1 , \bh_2+\bh_1+s_1-1 }(0,0).
	    \end{split}
	\end{equation} Note that the inclusion of the entire tower of ${\rm SL}(2, \mathbb{R})_R$ descendants is necessary to perform this non-local transformation in $\bz$ on the OPE.
	To obtain this expression, we use the integral identity
	\begin{equation}
	        \begin{split}
	            \int_{\mathbb{R}} \frac{d \bw}{2 \pi i} \bw^{n_1-1}(1-\bw)^{n_2-1}
	             = -\frac{\Gamma(1-n_1-n_2)}{\Gamma(1-n_1)\Gamma(1-n_2)},
	        \end{split}
	    \end{equation}
	    where the integral converges provided that ${\rm Re}(n_1)>0$, ${\rm Re}(n_2)>0$, and ${\rm Re}(n_1+n_2)<1$.  We analytically continue the result in $n_1$ and $n_2$, as in \cite{Sharma:2021gcz}. Note that for the application above, this amounts to an analytic continuation in $\bh_1$.
	    
	   Setting $(h_1, \bh_1)= (3-q, 1-q+\varepsilon)$ and $(h_2, \bh_2) = (h, \bh)$,  \eqref{lt-ope} becomes  
	   \begin{equation}
	    \begin{split}
	        {\bf L}[ \mathcal{O}_{3-q, 1-q+\varepsilon}] (z,\bz)\mathcal{O}_{h,\bh}(0,0) 
	              &\sim -\frac{\gamma^{2-\varepsilon,s}_{1-\varepsilon}}{z} \sum_{m = 0}^\infty    
	              \frac{B(3-2q+m+\varepsilon,2\bh+1-\varepsilon)\Gamma (2q-2-m-\varepsilon)}{m!\Gamma(-1-m+ \varepsilon)\Gamma (2q- 2\varepsilon)}\\& \quad \quad \quad \quad \quad \quad 
	              \quad  \quad\quad \quad \quad \quad  \times 
	             \bz^{2-2q+m+\varepsilon} 
	               \bar{\partial}^m     \mathcal{O}_{h+2-q , \bh+2-q }(0,0).
	    \end{split}
	\end{equation} 
	Notice that in the limit $\varepsilon\to 0$, the light transform  introduces a zero that cancels against the pole in the beta function from the original OPE coefficient. Taking this limit  and using  \eqref{eq:3ptcoupling} to identify the normalization with the gravitational coupling constant $\gamma^{2,s}_{1} = - \frac{1}{2} \kappa$, we obtain the following OPE between  ${\rm w}^q$ currents and matter fields: 
	 \begin{equation}\label{w-ope}
	    \begin{split}
	       {\rm w}^q(z, \bz)  \mathcal{O}_{h,\bh}(0,0) 
	              \sim 
	                  \frac{1}{2}\frac{1}{z} \sum_{m = 0}^{2q-3}  
	             \frac{(m+1)\Gamma(2\bh+1)}{\Gamma(2\bh+4-2q+m)}  
	             \bz^{2-2q+m} 
	               \bar{\partial}^m     \mathcal{O}_{h+2-q , \bh+2-q }(0,0).
	    \end{split}
	\end{equation} 
	In the above expression we have kept only the terms that are singular in $\bz$. 
	
	 Next, we introduce a conformally covariant mode expansion of ${\rm w}^q$: 
	\begin{equation}
	    {\rm w}^q(z, \bz)
	         = \sum_{m,n} \frac{{\rm w}^q_{m,n}}{z^{3-q+m}\bz^{q+n}}.
	\end{equation}
	If we treat $z$ and $\bz$ as independent complex variables, then we can extract modes by taking contour integrals of the form
	\begin{equation}
	     {\rm w}^q_{m, n}
	             =  \oint  \frac{dz}{2 \pi i} z^{2-q+m}
	                \oint  \frac{d\bz}{2 \pi i}
	                    \bz^{q+n-1} {\rm w}^q (z,\bz).
	\end{equation}
	The action of these modes on matter fields is then defined to be
	\begin{equation} \label{wmn-action}
	        \begin{split}
	            \left[ {\rm w}^q_{m,n}, \mathcal{O}_{h,\bh}(z,\bz) \right]
	            \equiv \oint_{z} \frac{dw}{2 \pi i} w^{2-q+m}
	                \oint_{\bz} \frac{d\bw}{2 \pi i}
	                    \bw^{q+n-1}  {\rm w}^q(w, \bw)
	                    \mathcal{O}_{h,\bh}(z,\bz). 
	        \end{split}
	  \end{equation} 
	 As before, we are ultimately interested in symmetry constraints on the leading term in a holomorphic limit of the OPE. Hence, we focus on transformations which do not mix ${\rm SL}(2, \mathbb{R})_L$  primaries and descendants. These are generated by modes with $m=q-2$,  denoted by 
	\begin{equation}
	    \widehat{\rm w}^q_n
	        \equiv {\rm w}^q_{q-2, n}, 
	        \quad \quad 1-q \leq n \leq q-1.
	\end{equation} 
	Here the range of $n$ is restricted to span the wedge subalgebra of ${\rm w}_{1+\infty}$.\footnote{We expect other modes of ${\rm w}^q$ to generate local enhancements of these symmetries. For example, notice that when $q = 2$, the contour integral in $z$ picks out the zero mode with respect to ${\rm SL}(2, \mathbb{R})_L$. The OPE
	   \begin{equation} \nonumber
	   \oint_0 \frac{dz}{2 \pi i}   {\rm w}^2(z, \bz)   \mathcal{O}_{h,\bh}(0,0) 
	            \sim   \frac{\bar{h} \mathcal{O}_{h,\bar{h}}(0,0)}{\bar{z}^2} + \frac{\bar{\partial} \mathcal{O}_{h,\bar{h}}(0,0)}{\bar{z}} 
	   \end{equation} resembles that of a local stress tensor whose modes generate a local Virasoro symmetry. It would be interesting to study the additional constraints implied by these local enhancements, such as in the recent analysis of Virasoro \cite{Banerjee:2021dlm}.} Using the OPE \eqref{w-ope} and the identity 
	    \begin{equation}
	       \oint_{\bz} \frac{d\bw}{2 \pi i}
	                    \bw^{q+n-1}
  (\bw-\bz)^{2-2q+m} 
    = {q+n-1 \choose 2q-3 -m}  \bz^{2-q+m+n}, 
	    \end{equation}
	    and re-indexing the sum by $\ell =2q-3-m$,  we find that the action of these charges on matter fields is given by 
	        \begin{equation} \label{wcharge-action}
	        \begin{split}
	            \left[ \widehat{\rm w}^q_n, \mathcal{O}_{h,\bh}(z,\bz) \right]
	            & =     \frac{1}{2}  \sum_{\ell = 0}^{2q-3}  
	            {q+n-1\choose \ell}
	            \frac{   (2q-2-\ell)  \Gamma (2 \bh+1)}{\Gamma (2\bh+1-\ell) } 
	           \bz^{q+n-1-\ell} 
	               \partial_{\bz}^{2q-3-\ell}    \mathcal{O}_{h+ 2-q , \bh + 2-q } (z,\bz).
	        \end{split}
	    \end{equation} 
	   This action respects the relation
	    \begin{equation} \label{text-id}
	        \begin{split}
	            \left[ \widehat{\rm w}^p_m, \left[ \widehat{\rm w}^q_n, \mathcal{O}_{h,\bh}(z,\bz)\right]\right]
	            -  \left[\widehat{\rm w}^q_n, \left[\widehat{\rm w}^p_m, \mathcal{O}_{h,\bh}(z,\bz)\right]\right]
	            = \left[\left[\widehat{\rm w}^p_m, \widehat{\rm w}^q_n\right],\mathcal{O}_{h,\bh}(z,\bz) \right],
	        \end{split}
	    \end{equation}
	    where
	    \begin{equation} \label{w-algebra}
	        \left[\widehat{\rm w}^p_m, \widehat{\rm w}^q_n\right]
	         = \left[m (q-1)-n(p-1)\right] \widehat{\rm w}^{p+q-2}_{m+n}.
	    \end{equation}
	    Thus, our charges generate the action of a ${\rm w}_{1+\infty}$ symmetry.\footnote{It would be interesting to derive  the commutation relation \eqref{w-algebra} directly from the ${\rm w}^p {\rm w}^q$ OPE by taking an appropriate set of contour integrals.  Also, there may exist a redefinition of the modes, such as the one in \cite{Strominger:2021lvk}, under which the action \eqref{wcharge-action} simplifies to a familiar representation of ${\rm w}_{1+\infty}$.} A proof of \eqref{text-id} is provided in Appendix \ref{app:w-proof}.    The proof for general $p$ and $m$ is performed by induction, using $\widehat{\rm w}^{2}_k$ to vary $m$ and $\widehat{\rm w}^{\frac{5}{2}}_k$ to raise $p$.\footnote{The structure of the proof is consistent with the following observations made in \cite{Guevara:2021abz,Strominger:2021lvk}.  First, $\widehat{\rm w}^2_k$ generates an ${\rm SL}(2, \mathbb{R})$ subalgebra under which  other $\widehat{\rm w}^q_n$ transform covariantly like the modes of an ${\rm SL}(2, \mathbb{R})$ primary of weight $q$.  Second, the modes with $p \leq 2$ form a closed subalgebra, while $\widehat{\rm w}^{\frac{5}{2}}_k$ generates all higher-weight modes with $p > \frac{5}{2}$.
	    
	   }

	  It is interesting to note that in a general effective field theory  at tree-level the symmetries generated by $\widehat{\rm w}^q_n$ with $q>2$ receive corrections  \cite{Elvang:2016qvq, Laddha:2017vfh}.  These arise from non-minimal three-point couplings  of gravitons to matter fields and therefore depend on the bulk three-point coupling constants $\gamma^{s_1,s_2}_p$ associated with these interactions.   The leading and subleading soft graviton theorems are uncorrected by three-point curvature couplings of gravitons to matter because these interactions involve enough extra powers of momentum to vanish in a soft expansion to first subleading order. Equivalently, our general formula \eqref{mainresult2} reveals that the OPE coefficients for these curvature couplings do not contain poles when the graviton dimension $\Delta$ is taken to $\Delta = 1, 0$. Thus the symmetries generated by  $\widehat{\rm w}^{\frac{3}{2}}_m$ and $\widehat{\rm w}^2_{m}$ respectively are uncorrected. For $q>2$, the corrections are controlled by a finite number of parameters. This is due to the fact that there are only a finite number of massless three-point interactions in theories with massless interacting particles of spin $\leq2$ and therefore only a finite number of bulk coupling constants $\gamma^{s_1,s_2}_p$ parametrizing the corrections.  We leave the deformation of the charges and symmetry algebra induced by these terms as well as by loops to a future investigation.

	    We close this section with several comments on the symmetry action \eqref{wcharge-action} for specific values of $q$.  First,  when $q = 2$,  notice that \eqref{wcharge-action} takes the form  \begin{equation} \label{eq:sl2r-w}
        \begin{split}
                  \left[\widehat{\rm w}^2_{m}, \mathcal{O}_{h, \bh}(z, \bz) \right]
                 &=\bz^m \left( (m+1) \bh+ \bz \partial_{\bz}\right)\mathcal{O}_{h, \bh}(z, \bz).
        \end{split} 
    \end{equation}
   This is \emph{precisely} the action of the ${\rm SL}(2, \mathbb{R})_R$ generators in \eqref{eq:sl2r-r}!  Therefore, we have identified  the ${\rm SL}(2, \mathbb{R})$ subalgebra of ${\rm w}_{1+\infty}$ generated by $\widehat{\rm w}^2_m$ with  ${\rm SL}(2, \mathbb{R})_R$
    \begin{equation}
        \widehat{\rm w}^2_m = \bar{L}_m.
    \end{equation}
   We also note that $\widehat{\rm w}^2_m$ properly generates ${\rm SL}(2, \mathbb{R})_R$ transformations on the other $\widehat{\rm w}^q_n$ charges.  This can be seen from \eqref{w-algebra} by setting $p = 2$ and  using \eqref{w-weight} suggestively to replace $q$ with $\bh$, so that \eqref{w-algebra} takes the form 
   \begin{equation}
        \left[\widehat{\rm w}^2_m, \widehat{\rm w}^\bh_n\right]
            =\left(m (\bh-1)-n \right) \widehat{\rm w}^{\bh}_{m+n}.
   \end{equation}
  This is just the ${\rm SL}(2, \mathbb{R})_R$ transformation of the $n$th mode of a weight $\bh$ primary.   We therefore conclude that there are a finite number of modes  $1-q \leq n \leq q - 1$ for each ${\rm w}^q$ that form a  $(2q-1)$-dimensional closed algebra with the ${\rm SL}(2, \mathbb{R})_R$ generators.  These modes form the wedge subalgebra of ${\rm w}_{1+\infty}$.  It would be interesting to study the implications of this result on the allowed vacuum structure of the  celestial CFT.  We leave this question to future work.

   Next, observe that when $q = \frac{3}{2}$, \eqref{wcharge-action} becomes  
   \begin{equation}
     \begin{split}
       \left[\widehat{\rm w}^{\frac{3}{2}}_{n}, \mathcal{O}_{h, \bh}(z, \bz) \right]
                 &= \frac{1}{2}\bz^{n+ \frac{1}{2}} \mathcal{O}_{h+ \frac{1}{2}, \bh+ \frac{1}{2}}(z, \bz).
    \end{split}
   \end{equation}
   In fact, using \eqref{w-ope} and  \eqref{wmn-action}, we find that the $m, n = \pm \frac{1}{2}$ modes of ${\rm w}^{\frac{3}{2}}$ generate the transformation
   \begin{equation}
       \left[{\rm w}^{\frac{3}{2}}_{m,n},\mathcal{O}_{h, \bh}(z, \bz) \right]
        = \frac{1}{2}z^{m+ \frac{1}{2}}\bz^{n+ \frac{1}{2}} \mathcal{O}_{h+ \frac{1}{2}, \bh+ \frac{1}{2}}(z, \bz),
   \end{equation}
  which can be identified with the action of the translation generators in \eqref{eq:translations}.  Specifically, we can further identify
   \begin{equation}
        {\rm w}^{\frac{3}{2}}_{m,n} = P_{m,n}.
   \end{equation}
   Note that these are modes of ${\rm w}^{\frac{3}{2}}$, which is a $\Delta = 3$ scalar operator.   Our work provides a simple construction of this operator, the existence of which was previously discussed in \cite{Barnich:2017ubf,Fotopoulos:2019vac,Fotopoulos:2020bqj,Pasterski:2021fjn}.

 \section{${\rm w}_{1+\infty}$ Symmetry of the Celestial OPE}
    \label{sec:w-ope}
    
    In this section, we show that the OPE coefficients \eqref{mainresult2} derived from Poincar\'e in Section \ref{sec:ope_sym} and verified from momentum space in Section \ref{sec:OPEcol} respect the action generated by $\widehat{\rm w}^q_n$.  Here, we focus only on the constraints generated by positive-helicity gravitons and leave those associated to negative-helicity  gravitons to future work.   In particular, we study the transformations derived in the previous section from minimally coupled gravitons. In doing so, we assume that the finite number of effective field theory corrections (including those enumerated in\cite{Elvang:2016qvq}) are absent.

    Consider the transformation generated by the top component with $n = q-1$.  Using \eqref{wcharge-action}, we find the action on primaries takes the form  \begin{equation} \label{wtop-primary}
	        \begin{split}
	            \left[\widehat{\rm w}^q_{q-1}, \mathcal{O}_{h,\bh}(z,\bz) \right]
	            & =  (q-1)     \sum_{\ell = 0}^{2q-3}  
	            {2q-3\choose \ell}
	            \frac{ \Gamma (2 \bh+1)}{\Gamma (2\bh+1-\ell) } 
	           \bz^{2q-2-\ell} 
	               \partial_{\bz}^{2q-3-\ell}    \mathcal{O}_{h+ 2-q , \bh + 2-q } (z,\bz).
	        \end{split}
	    \end{equation} 
	  The action on a descendant is then given by
	  \begin{equation}
	      \begin{split}
	          \left[\widehat{\rm w}^q_{q-1}, \partial_{\bz}^m \mathcal{O}_{h,\bh}(z,\bz) \right]
	             &=\partial_{\bz}^m \left[\widehat{\rm w}^q_{q-1}, \mathcal{O}_{h,\bh}(z,\bz) \right]\\
	            & =
	             (q-1)     \sum_{\ell = 0}^{2q-3}  
	            {2q-3\choose \ell}
	            \frac{ \Gamma (2 \bh+1)}{\Gamma (2\bh+1-\ell) } 
	      \partial_{\bz}^m \left( \bz^{2q-2-\ell} 
	               \partial_{\bz}^{2q-3-\ell}    \mathcal{O}_{h+ 2-q , \bh + 2-q } (z,\bz)\right), 
	      \end{split}
	  \end{equation}
	  which, upon placing the operator  $\mathcal{O}_{h,\bh}$ at the origin, simplifies to 
	  \begin{equation} \label{w-desc}
	    \begin{split}
	     \left[\widehat{\rm w}^q_{q-1}, \bar{\partial}^m\mathcal{O}_{h,\bh}(0,0) \right]
	        =\frac{m(q-1) \Gamma (2 \bh+m)}{\Gamma (2 \bh+m-2 q+3)} \bar{\partial}^{m-1}     \mathcal{O}_{h+ 2-q , \bh + 2-q } (0,0)
	            .
	    \end{split}
	  \end{equation}
	  
	   Now, acting with $\widehat{\rm w}^q_{q-1}$ on the left-hand side of \eqref{ansatz}, we find 
	  \begin{equation}   \label{w-leftaction}
	    \begin{split}
	      & \left[\widehat{\rm w}^q_{q-1}, \mathcal{O}_{h_1, \bh_1}(z,\bz)\mathcal{O}_{h_2,\bh_2}(0,0)\right]\\
	            & \quad \quad  =   (q-1)     \sum_{\ell = 0}^{2q-3}  
	            {2q-3\choose \ell}
	            \frac{ \Gamma (2 \bh_1+1)}{\Gamma (2\bh_1+1-\ell) } 
	           \bz^{2q-2-\ell} 
	               \partial_{\bz}^{2q-3-\ell}      \mathcal{O}_{h_1+ 2-q , \bh_1 + 2-q } (z,\bz)\mathcal{O}_{h_2,\bh_2}(0,0)
	                \\
	           & \quad \quad \sim  
	           \frac{\bz^{p+1}}{z}\sum_{m = 0}^\infty    C^{(m)}_{p} ( \bh_1{+}2{-}q,  \bh_2) 
	                    \frac{(q-1)\Gamma (2 \bh_1+m+p+1)}{\Gamma (2 \bh_1+m+p-2 q+4)}
	                    \bz^{ m} 
	               \bar{\partial}^m   \mathcal{O}_{h_1+h_2-q+1, \bh_1+\bh_2+p-q+2}(0,0). 
	    \end{split}
	\end{equation}
	Using \eqref{w-desc} to act with $\widehat{\rm w}^q_{q-1}$ on the right-hand side of \eqref{ansatz}  gives
	\begin{equation}
	    \begin{split}
	        \frac{1}{z}\sum_{m = 0}^\infty& C^{(m)}_{p} ( \bh_1,  \bh_2)     \bz^{p+m} 
	             \left[ \widehat{\rm w}^q_{q-1} ,\bar{\partial}^m   \mathcal{O}_{h_1+h_2-1, \bh_1+\bh_2+p}(0,0)\right]\\
	       &= \frac{ \bz^{p+1}}{z}\sum_{m = 0}^\infty C^{(m+1)}_{p} ( \bh_1,  \bh_2)  \frac{(m+1)(q-1) \Gamma (2 \bh_1+2\bh_2+2p+m+1)}{\Gamma (2 \bh_1+2\bh_2+2p+m-2q+4)}\\&
	             \quad \quad \quad
	               \quad \quad \quad  \quad \quad \quad  \quad \quad \quad
	                 \quad   \quad \quad \quad  \quad \quad \quad
	                 \times 
	                 \bz^{m}   \bar{\partial}^{m}     \mathcal{O}_{h_1+h_2-q+1 , \bh_1+\bh_2+p-q + 2 } (0,0) ,
	    \end{split}
	\end{equation}
	and equating the result to \eqref{w-leftaction}, 
	we obtain the constraint
	\begin{equation} \label{w-constraint}
	   \begin{split}
	         \frac{ \Gamma (2 \bh_1+m+p+1)}{\Gamma (2 \bh_1+m+p-2 q+4)}
	         &  C^{(m)}_{p} ( \bh_1+2-q,  \bh_2) 
	               = \frac{(m+1)  \Gamma (2 \bh_1+2\bh_2+2p +m+1)}{\Gamma (2 \bh_1+2\bh_2+2p+m-2q+4)}  C^{(m+1)}_{p} ( \bh_1,  \bh_2).
	   \end{split}
	\end{equation}
	 It is straightforward to  show that our solution \eqref{mainresult2} satisfies this constraint. 
	 
	 In deriving this constraint, we have assumed the absence of the finite number of three-point curvature couplings to matter that correct the action \eqref{wcharge-action}, including those enumerated in\cite{Elvang:2016qvq}.
	 In the presence of these corrections, the analysis we present in this section must be generalized,\footnote{A naive accounting of these terms suggests that they must all cancel against one another to give \eqref{w-constraint}. In order to do so, the three-point couplings $\gamma^{s_1 ,s_2}_p$ must all be $\co(1)$ in dimensionless units, and therefore consistent with an effective field theoretic expectation of naturalness.} yet  must still  ultimately yield a constraint that is solved by the Poincar\'e-consistent solution \eqref{mainresult2}. We leave a systematic understanding of these corrections to future work.

	\section*{Acknowledgements}
		
 		We are grateful to Alfredo Guevara, Hofie Hannesdottir, Jakob Salzer, and Andy Strominger for insightful conversations. This work was  supported by DOE grant de-sc/000787 and the Black Hole Initiative at Harvard University, which is funded by grants from the John Templeton Foundation and the Gordon and Betty Moore Foundation.	M.P.  also receives support from the Harvard Society of Fellows.   
		
	\begin{appendix}
	
		\section{Conventions}
		\label{appA}
		
		Null four-momenta are parametrized in the following way: 
		\be \label{mom-par}
			p^\mu_i= \frac{\eps_i \omega_i}{\sqrt{2}} \big(1+ z_i \bz_i, z_i+ \bz_i, -i (z_i- \bz_i) ,  1- z_i \bz_i\big).
		\ee 
		Here $(z_i, \bz_i)$ specify a point on the celestial plane, $\omega_i$ specifies an energy and $\epsilon_i = \pm 1$ for  outgoing/incoming particles. These obey
		\be
			p_i \cdot p_j = -\eps_i \eps_j \omega_i \omega_j z_{ij} \bz_{ij}.
		\ee
		We introduce
		 \be
		 	\sigma^\mu = (1,  \sigma^i ), \quad \quad \quad \bar \sigma^\mu =  (1, -\sigma^i),
		 \ee
		 where $\sigma^i$ are the standard Pauli matrices and
		 \be
		 	\gamma^\mu =  \left(\begin{matrix} 0 & (\sigma^\mu)_{\dot a  b} \\ (\bar \sigma^\mu)^{ a \dot b} &0\end {matrix} \right) .
		 \ee
		 Here, $\gamma^\mu$ are the usual gamma matrices obeying the Clifford algebra
		 \be
		 	\left \{ \gamma^\mu, \gamma^\nu\right\}= -2 \eta^{\mu\nu}.
		 \ee
		Defining
		  \be
		 	\begin{split}
		 	p_{\dot a  b}& \equiv \frac{1}{\sqrt{2}}p_\mu(\sigma^\mu)_{\dot a  b} =\frac{1}{\sqrt{2}} \left (\begin{matrix} -p^0 + p^3 & p^1 -i p^2\\ p^1 + i p^2 & -p^0 -p^3
					\end{matrix} \right)
					=- \eps \omega \left(\begin{matrix} z \bz&- \bz \\ -z &1
					\end{matrix}\right) , \\ 
		 	 p^{ a \dot b}& \equiv  \frac{1}{\sqrt{2}}p_\mu  (\bar \sigma^\mu)^{ a \dot b}  = - \frac{1}{\sqrt{2}} \left (\begin{matrix} p^0 + p^3 & p^1 -i p^2\\ p^1 + i p^2 & p^0 -p^3 
			\end{matrix} \right)
			=- \eps \omega \left(\begin{matrix} 1&\bz \\ z &z \bz
					\end{matrix}\right) ,
			\end{split}
		 \ee
		 we obtain expressions for null momenta in terms of two-component spinor helicity variables: 
		  \begin{equation}
		    \begin{split}
		 	p_{\dot a b} &= - | p]_{\dot a } \langle p |_b = -\tilde{\lambda}_{\dot a} \lambda_{b},\\ p^{a \dot b} &= - |p \rangle^{a} [p|^{\dot b}=-\lambda^{a}\tilde{\lambda}^{\dot b},
		 	\end{split}
		 \end{equation}
		 where
		 \be
		 	\begin{split}
				| p]_{\dot a } &= \sqrt{\omega}\left (\begin{matrix}- \bz \\1 \end{matrix}\right), \quad \quad \quad 
				 \langle p |_a = \eps \sqrt{\omega} \left(\begin{matrix}-z&1 \end{matrix}\right),\\
				 | p\rangle^{a} &= \eps \sqrt{\omega} 	\left (\begin{matrix} 1 \\z \end{matrix}\right), \quad \quad \quad ~~ 
				[ p |^{ \dot a} =  \sqrt{\omega} \left(\begin{matrix}1&\bz \end{matrix}\right).
			\end{split}
		 \ee 
		  Indices are raised and lowered with the Levi-Civita symbol: 
		 \be
		 	[p|^{\dot a} = \eps^{\dot a \dot b} |p]_{\dot b}, \quad \quad \quad |p\rangle^a = \eps^{ab} \langle p |_b
		 \ee
		 where
		 \be
		 	\eps^{ab} = \eps^{\dot a \dot b} = \left(\begin{matrix} 0&1 \\ -1&0\end{matrix}\right).
		 \ee 
		 The following inner products can be calculated by  matrix multiplication:
		  \be
		 	\begin{split}
		 	\langle ij \rangle &= \langle i|_a |j \rangle^a 
				=\eps_i \eps_j \sqrt{\omega_i \omega_j} \left(\begin{matrix}-z_i&1 \end{matrix}\right)\left (\begin{matrix} 1 \\z_j \end{matrix}\right)
				 = - \eps_i \eps_j \sqrt{\omega_i \omega_j}z_{ij}, \\
			[ ij ] &=[ i|^{\dot a} |j ]_{\dot a} 
				= \sqrt{\omega_i \omega_j} \left(\begin{matrix}1&\bz_i \end{matrix}\right)\left (\begin{matrix} -\bz_j \\ 1 \end{matrix}\right)
				 =   \sqrt{\omega_i \omega_j} \bz_{ij}.
			\end{split}
		 \ee 
		 Finally, in these conventions
		 \begin{equation}
		     p_i \cdot p_j = \langle ij \rangle[ ij ] ,
		 \end{equation}
		 which can be readily verified using the identity
		 \begin{equation}
		     \sigma^\mu_{\dot a a}~\bar \sigma^\nu{}^{ a \dot a} = -2 \eta^{\mu \nu}.
		 \end{equation}

	\section{Holographic Symmetry Algebra}
	    \label{app:old-sym}

	     Momentum space soft theorems imply that low-energy modes of gravitons generate asymptotic symmetries (see \cite{Strominger:2017zoo} and references therein). In boost weight space, these symmetry generators can be identified as gravitons of integer conformal weight $\Delta = 2, 1, 0, -1, \cdots$ and are referred to as conformally soft currents \cite{Puhm:2019zbl,Guevara:2019ypd,Adamo:2019ipt,Pate:2019lpp}.   In \cite{Guevara:2021abz}, positive-helicity graviton currents were found to organize into finite $(3-\Delta)$-dimensional ${\rm SL}(2, \mathbb{R})_R$ representations with $\bh = \frac{1}{2} (\Delta -2)$ and highest (lowest) weights $\frac{2-\Delta}{2}$ ($\frac{\Delta-2}{2}$).   An  alternate and ultimately equivalent approach to that in Section \ref{sec:w-gen} and \ref{sec:w-ope} is to perform an ${\rm SL}(2, \mathbb{R})_R$ mode expansion of these conformally soft currents and study the  transformations generated by the modes.

     We begin with a positive-helicity graviton  current defined by 
    \begin{equation}
        H^{k}(z, \bz) = \lim_{\varepsilon \to 0} \varepsilon G_{k + \varepsilon}^{+}(z, \bz) , \quad \quad \quad k = 2,1, 0, -1, -2, \ldots,
    \end{equation}
    where $G^{+}_{\Delta}(z, \bz)$ denotes a graviton of   boost weight $\Delta$ constructed by Mellin-transforming a positive-helicity graviton with respect to  energy. In the limit $\varepsilon\to 0$, the additional factor of $\varepsilon$ is needed to cancel the pole in the beta function from the original OPE coefficient. This factor serves the same purpose as the zero introduced by the light transform \eqref{wcurrent-def} in Section \ref{sec:w-gen}.  
    
    The current $H^k$ has left and right conformal weight
    \begin{equation} \label{H-weight}
       (h,\bar h)= \left(\frac{k+2}{2},  \frac{k-2}{2}\right),
    \end{equation}
    and admits a mode expansion \cite{Guevara:2021abz}
    \begin{equation}  \label{H-exp}
            \begin{split}
                H^{k}(z,\bar{z}) &= \sum_{n = \frac{k-2}{2}}^{\frac{2-k}{2}} \frac{H^{k}_{n}(z)}{\bar{z}^{n + \frac{k-2}{2}}}. 
            \end{split}
    \end{equation} 
    The modes $H^k_n$ are labelled by their transformation under ${\rm SL}(2, \mathbb{R})_R$.  Accordingly, they respect the following commutation relations with the ${\rm SL}(2, \mathbb{R})_R$ generators:
     \begin{equation}
         \left[\bar{L}_m, H^k_n\right] = \left(\frac{k-4}{2}m -n\right) H^k_{m+n}.
     \end{equation}
     Note that because of the truncated mode expansion \eqref{H-exp}, the mode number $n$ on $H^k_n$ cannot be lowered indefinitely by acting with $\bar{L}_{-1}$. $H^k_{\frac{k-4}{2}}$ is both a primary and a descendant, and therefore null in an appropriately-defined ${\rm SL}(2,\mathbb{R})_R$ covariant norm for celestial conformal field theory \cite{Guevara:2021abz,Pasterski:2021fjn,Crawley:2021ivb}.\footnote{The importance of global conformal primary descendants of conformally soft celestial operators was first pointed out in \cite{Banerjee:2019aoy,Banerjee:2019tam}.} This is in contrast with the $\widehat{\rm w}^q_n$ modes, which form a closed algebra with the ${\rm SL} (2, \mathbb{R})_R$ generators without any primary descendants or null states.

    Using the minimal-coupling graviton OPE \eqref{minimal-ope} with $\bh_1 = \frac{k+\varepsilon}{2}-1$ and $s_1 = 2$, we find that the OPE of $H^k$ with matter is given by 
    \begin{equation}
       H^k (z_1, \bz_1) \mathcal{O}_{h, \bh} (z_2, \bz_2)
					\sim  -\frac{\kappa}{2} \frac{1}{z_{12}} \sum_{m = 0}^{1-k}  
						  \frac{(-1)^{k+m-1}}{m!(1-k-m)!} \frac{\Gamma( 2\bh+1)}{\Gamma(2\bh+k  +m )} 
							 \bz_{12}^{m+1} \partial_{\bz_2}^m \mathcal{O}_{h+\frac{1}{2} k ,  \bh+ \frac{1}{2} k}(z_2, \bz_2).
    \end{equation}
    Replacing $H^k$ with its mode expansion given in  \eqref{H-exp}  and matching powers of $\bz_1$, the above OPE implies 
    \begin{equation}
        \begin{split} \label{Hkn-OPE}
           H^{k}_{n}(z_1)  \mathcal{O}_{h, \bh} (z_2, \bz_2)
			= \frac{\kappa}{2} \frac{1}{z_{12}} \sum_{m = 0}^{1-k}  
						  \frac{(-1)^{-\frac{1}{2}k+n}(m+1)}{(1-k-m)!} \frac{\Gamma( 2\bh+1)}{\Gamma(2\bh+k  +m )} 
						  \frac{\bz_2^{m+\frac{k}{2}+n} \partial_{\bz_2}^m \mathcal{O}_{h+\frac{k}{2}  ,  \bh+ \frac{k}{2} }(z_2, \bz_2)}{(\frac{2-k}{2}-n)!(m+\frac{k}{2}+n)!} .
        \end{split}
    \end{equation} 
    Introducing charges  that act on operators via
    \begin{equation}
        \left[\widehat H^k_m, \co_{h, \bh} (z, \bz)\right]
             =  \oint_{z}  \frac{dw}{2 \pi i} H^{k}_{m}(w)
                \co_{h, \bh} (z, \bz),
    \end{equation}
    we obtain the following transformations of celestial primary operators:\footnote{Note that the $k =1,0,-1$ transformations appeared previously in \cite{Banerjee:2020zlg,Banerjee:2021cly}.  Our modes are labelled according to ${\rm SL}(2,\mathbb{R})_R$ and are related to the modes $P_{m,n}$, $J^a_n$ in \cite{Banerjee:2020zlg} and $S_n^m$ in \cite{Banerjee:2021cly} by   $\widehat H^{1}_{\frac{1}{2} + m} = (-1)^{m}P_{0,m}$, $\widehat H^{0}_{m} = \frac{2 (-1)^m}{(1+m)!(1-m)!}J^m_0$, and $\widehat H^{-1}_{\frac{3}{2} - m}=S_0^m$. }
    \begin{equation}
        \begin{split}
            \left[\widehat H^k_{n}, \mathcal{O}_{h, \bh} (z, \bz)\right]
			= \frac{\kappa}{2}   \sum_{m = 0}^{1-k}  
						  \frac{(-1)^{-\frac{1}{2}k+n}(m+1)}{(1-k-m)!} \frac{\Gamma( 2\bh+1)}{\Gamma(2\bh+k  +m )} 
						  \frac{\bz^{m+\frac{k}{2}+n} \partial_{\bz}^m \mathcal{O}_{h+\frac{k}{2}  ,  \bh+ \frac{k}{2} }(z, \bz)}{(\frac{2-k}{2}-n)!(m+\frac{k}{2}+n)!}.
        \end{split}
    \end{equation}
    Comparing this with the action of the $\widehat{\rm w}^q_n$ modes \eqref{wcharge-action}, we find
    \begin{equation}
        \widehat{\rm w}^q_n =(-1)^{n+q} \frac{1}{\kappa}\Gamma(q+n)\Gamma (q-n)
           \widehat H^{4-2q}_n.
    \end{equation}
     Up to the sign $(-1)^{q+n}$, this is precisely the relation found in \cite{Strominger:2021lvk} between the chiral currents $H^{-2p+4}_m$ from \cite{Guevara:2021abz} and chiral currents $w^p_m$ that generate a ${\rm w}_{1+\infty}$ algebra.  $w^p_m \to (-1)^{p+m} w^p_m$ is the composition of two automorphisms of the ${\rm w}_{1+\infty}$ algebra: an inner automorphism $w^p_m \to (-1)^m w^p_m$ (given by conjugation by $e^{i\pi w^2_0}$) and $w^p_m \to (-1)^{p} w^p_m$.  Therefore, the algebra is unaffected by this relative sign and the modes $\widehat H^k_n$ obey the algebra found in \cite{Guevara:2021abz}.

     Finally, one can follow the same logic as presented in Section \ref{sec:w-ope}  using $\widehat H^k_n$ instead of  $\widehat{\rm w}^q_n$.  The analysis is exactly the same because up to a constant of proportionality, the action on primaries of the top component $\widehat H^k_{\frac{2-k}{2}}$  is equal to that of the top component $\widehat{\rm w}^q_{q-1}$ for $k= 4-2q$.   This also implies that they have the same action on descendants, which is given by
	  \begin{equation}
	        \left[\widehat H^k_{\frac{2-k}{2}}, \partial_\bz^m\co_{h, \bh}(z,\bz) \right]
	            =\partial_\bz^m\left[\widehat H^k_{\frac{2-k}{2}}, \co_{h, \bh}(z,\bz) \right].
	  \end{equation}  
	  Note that if one works directly with the modes of ${\rm w}^q$ and $H^k$ and their  commutation relations with the global conformal generators, then intermediate expressions in the two analyses differ  due to the difference in right-moving weights.  However, these differences at intermediate steps ultimately cancel to yield the same result.

\section{${\rm w}_{1+\infty}$ Commutators}
\label{app:w-proof}

In this appendix, we prove that the action  \eqref{wcharge-action} obeys 
 \begin{equation} \label{jacobi-id}
	        \begin{split}
	            \left[\widehat{\rm w}^p_m, \left[\widehat{\rm w}^q_n, \mathcal{O}_{h,\bh} \right]\right]
	            -  \left[\widehat{\rm w}^q_n, \left[\widehat{\rm w}^p_m, \mathcal{O}_{h,\bh} \right]\right]
	            =  \left(m (q-1)-n(p-1)\right)\left[\widehat{\rm w}^{p+q-2}_{m+n},\mathcal{O}_{h,\bh} \right].
	        \end{split}
	    \end{equation}
Here we denote the action \eqref{wcharge-action}  by 
 \begin{equation} \label{wcharge-action-app}
	        \begin{split}
	            \left[\widehat{\rm w}^q_n, \mathcal{O}_{h,\bh}  \right]
	            & =  \mathcal{D}^q_n (\bh)       \mathcal{O}_{h+ 2-q , \bh + 2-q } ,
	        \end{split}
\end{equation}
where $\mathcal{D}^q_n (\bh)$ is the differential operator
 \begin{equation} \label{w-diff}
	        \begin{split}
	         \mathcal{D}^q_n (\bh) &\equiv   \frac{1}{2}  \sum_{\ell = 0}^{2q-3}  
	            {q+n-1\choose \ell}
	            \frac{   (2q-2-\ell)  \Gamma (2 \bh+1)}{\Gamma (2\bh+1-\ell) } 
	           \bz^{q+n-1-\ell} 
	               \partial_{\bz}^{2q-3-\ell} 
	        \end{split}
	    \end{equation}
	 and the position of the operator is suppressed to simplify notation.
	 
	 An outline of the proof is as follows.  In Subsection \ref{app:special}, we prove \eqref{jacobi-id} for all $q\geq1$ and $n \in [1-q, q-1]$ and $(p,m) = (1,0)$, $(\frac{3}{2}, -\frac{1}{2})$, and $(\frac{5}{2},-\frac{3}{2})$.  Then, in Subsection \ref{app:allm}, we use the action of $\widehat{\rm w}^2_k$ to prove that if \eqref{jacobi-id} holds  for all $q\geq1$ and $n \in [1-q, q-1]$ and  any given $p$ and $m \in [1-p, p-1]$, it holds for  that given $p$ for all $m\in [1-p, p-1]$.  Note that this result follows from the fact that, by construction, the modes transform covariantly under ${\rm SL}(2, \mathbb{R})_R$.  Nevertheless, we present the argument because it is instructive for the subsequent subsection.  In the last subsection, we use the action of $\widehat{\rm w}^{\frac{5}{2}}_k$ to prove that if \eqref{jacobi-id} holds for a given $p$ and all $m \in [1-p, p-1]$, $q$ and $n\in [1-q, q-1]$, then it holds for $p +\frac{1}{2}$ and all $m\in [\frac{1}{2}-p,p-\frac{1}{2}]$, provided $(p,m) \neq (1,0)$. Thus, starting with the base case of $(p,m) = (\frac{3}{2}, -\frac{1}{2})$,  we conclude that \eqref{jacobi-id} holds for all $p,q\geq 1$, $m \in [1-p,p-1]$, and $n\in [1-q,q-1]$.
 
	   \addtocontents{toc}{\protect\setcounter{tocdepth}{1}}
	   \subsection{Special Cases}
	        \label{app:special}

        In this subsection, we prove  \eqref{jacobi-id}  by brute force for all values of $q\geq 1$ and $n \in [1-q, q-1]$ and select choices of $p$ and $m$.
        
        First consider $\widehat{\rm w}^1_0$, the only nonzero component of $\widehat{\rm w}^1_m$.  In this case, \eqref{w-algebra} and  \eqref{wcharge-action-app} reduce to 
        \begin{equation} \label{w10}
            \left[\widehat{\rm w}^1_0, \widehat{\rm w}^q_n \right] = 0, \quad \quad \quad 
            \left[\widehat{\rm w}^1_0, \mathcal{O}_{h, \bh}\right] =0.
        \end{equation}
        Note that by definition, the symmetry action \eqref{wmn-action} is only non-trivial when the ${\rm w}^q(z,\bz) \mathcal{O}_{h,\bh}(0,0)$ OPE has singularities in both $z$ and $\bz$. This definition implies the latter equation in  \eqref{w10}.  A non-trivial $\widehat{\rm w}^1_0$  action could be possible using a different definition of symmetry action that is non-trivial for OPEs with only singularities in $z$.
        Then, noting that 
        \begin{equation}
            \left[\widehat{\rm w}^1_0,\left[ \widehat{\rm w}^q_n , \mathcal{O}_{h, \bh}\right]\right]
                 = \mathcal{D}^q_n(\bh)
                    \left[\widehat{\rm w}^1_0,\mathcal{O}_{h+2-q, \bh+2-q}\right] = 0,
        \end{equation}
        we find \eqref{jacobi-id} holds for $(p,m)=(1,0)$ and  all $(q,n)$. 
        
        Next, when $(p,m) = (\frac{3}{2},-\frac{1}{2})$, we have
    \begin{equation}
    \begin{aligned}
    \left[\widehat{\rm w}_{-\frac{1}{2}}^{\frac{3}{2}},\left[\widehat{\rm w}_n^q,\mathcal{O}_{h,\bar{h}}\right]\right] &- \left[\widehat{\rm w}_n^q,\left[\widehat{\rm w}_{-\frac{1}{2}}^{\frac{3}{2}},\mathcal{O}_{h,\bar{h}}\right]\right] \\
    &=  \left[\mathcal{D}_n^q(\bar{h})\mathcal{D}_{-\frac{1}{2}}^{\frac{3}{2}}(\bar{h}+2-q)  - \mathcal{D}_{-\frac{1}{2}}^{\frac{3}{2}}(\bar{h})\mathcal{D}_n^q\bigg(\bar{h}+\frac{1}{2}\bigg)\right]\mathcal{O}_{h+\frac{5}{2}-q,\bar{h}+\frac{5}{2}-q},
    \end{aligned}
    \end{equation}
    which is written explicitly as 
    \begin{equation}
    \left(\frac{1}{2}\right)^2 \sum_{\ell=0}^{2q-3} {q+n-1 \choose \ell} \frac{(2q-2-\ell)(-\ell)\Gamma(2\bar{h}+1)}{\Gamma(2\bar{h} + 2 - \ell)}\bar{z}^{q+n-1-\ell}\partial_{\bar{z}}^{2q-3-\ell}\mathcal{O}_{h+\frac{5}{2}-q,\bar{h}+\frac{5}{2}-q}. 
    \end{equation}
    Upon re-indexing the sum, this becomes
    \begin{equation}
    \begin{aligned}
    -\left(\frac{1}{2}\right)^2 \sum_{\ell=0}^{2q-4}(q+n-1) {q+n-2 \choose \ell} &\frac{(2q-3-\ell)\Gamma(2\bar{h}+1)}{\Gamma(2\bar{h} + 1 - \ell)}\bar{z}^{q+n-2-\ell}\partial_{\bar{z}}^{2q-4-\ell}\mathcal{O}_{h+\frac{5}{2}-q,\bar{h}+\frac{5}{2}-q} \\
    &= - \frac{1}{2} (q+n-1) \mathcal{D}_{n-\frac{1}{2}}^{q-\frac{1}{2}}(\bar{h})\mathcal{O}_{h+\frac{5}{2}-q,\bar{h}+\frac{5}{2}-q},
    \end{aligned}
    \end{equation} 
    which demonstrates that 
    \begin{equation}
    \left[\widehat{\rm w}_{-\frac{1}{2}}^{\frac{3}{2}},\left[\widehat{\rm w}_n^q,\mathcal{O}_{h,\bar{h}}\right]\right] - \left[\widehat{\rm w}_n^q,\left[\widehat{\rm w}_{-\frac{1}{2}}^{\frac{3}{2}},\mathcal{O}_{h,\bar{h}}\right]\right] = - \frac{1}{2} (q+n-1) \left[\widehat{\rm w}_{n-\frac{1}{2}}^{q-\frac{1}{2}},\mathcal{O}_{h,\bar{h}}\right].           
    \end{equation}
    Similarly, for  $(p,m) = (\frac{5}{2},-\frac{3}{2})$, we compute
\begin{equation}
\begin{aligned}
\left[\widehat{\rm w}_{-\frac{3}{2}}^{\frac{5}{2}},\left[\widehat{\rm w}_n^q,\mathcal{O}_{h,\bar{h}}\right]\right] &- \left[\widehat{\rm w}_n^q,\left[\widehat{\rm w}_{-\frac{3}{2}}^{\frac{5}{2}},\mathcal{O}_{h,\bar{h}}\right]\right] \\  
&=  \left[\mathcal{D}_n^q(\bar{h})\mathcal{D}_{-\frac{3}{2}}^{\frac{5}{2}}(\bar{h}+2-q)  - \mathcal{D}_{-\frac{3}{2}}^{\frac{5}{2}}(\bar{h})\mathcal{D}_n^q\bigg(\bar{h}-\frac{1}{2}\bigg)\right]\mathcal{O}_{h+\frac{3}{2}-q,\bar{h}+\frac{3}{2}-q}.  \\
\end{aligned}
\end{equation}
Explicitly, this is the sum 
\begin{equation}
\begin{aligned}
 \frac{3}{4}\sum_{\ell=0}^{2q-3}{q {+} n {-}1 \choose \ell}(2q{-}2{-}\ell)\bigg( &\frac{\ell \ \Gamma(2\bar{h})}{\Gamma(2\bar{h} {+} 1 {-} \ell)}\bar{z}^{q+n-1-\ell}\partial_{\bar{z}}^{2q-1-\ell} - \frac{2(q{+}n{-}1{-}\ell)\Gamma(2\bar{h})}{\Gamma(2\bar{h}{-}\ell)}\bar{z}^{q+n-2-\ell}\partial_{\bar{z}}^{2q-2-\ell} \\
&\qquad - \frac{(q{+}n{-}1{-}\ell)(q{+}n{-}2{-}\ell)\Gamma(2\bar{h})}{\Gamma(2\bar{h}{-}\ell)}\bar{z}^{q+n-3-\ell}\partial_{\bar{z}}^{2q-3-\ell} \bigg)\mathcal{O}_{h+\frac{3}{2}-q,\bar{h}+\frac{3}{2}-q},
\end{aligned}
\end{equation}
which upon appropriately re-indexing each term and simplifying becomes  
\begin{equation}
  \begin{aligned}
    -\frac{3}{4}\sum_{\ell=0}^{2q-2}(q+n-1){q+n-2\choose \ell}& \frac{(2q-1-\ell)\Gamma(2\bh+1)}{\Gamma(2\bh+1-\ell)} \bar{z}^{q+n-2-\ell}\partial_{\bar{z}}^{2q-2-\ell} \mathcal{O}_{h+\frac{3}{2}-q,\bar{h}+\frac{3}{2}-q}\\
    & \quad \quad \quad \quad \quad \quad = -\frac{3}{2} (q+n-1) \mathcal{D}_{n-\frac{3}{2}}^{q+\frac{1}{2}}(\bar{h})\mathcal{O}_{h+\frac{3}{2}-q,\bar{h}+\frac{3}{2}-q} . 
  \end{aligned}
\end{equation} 
Thus we have demonstrated that 
    \begin{equation} \label{eq:w52}
    \left[\widehat{\rm w}_{-\frac{3}{2}}^{\frac{5}{2}},\left[\widehat{\rm w}_n^q,\mathcal{O}_{h,\bar{h}}\right]\right] - \left[\widehat{\rm w}_n^q,\left[\widehat{\rm w}_{-\frac{3}{2}}^{\frac{5}{2}},\mathcal{O}_{h,\bar{h}}\right]\right] = - \frac{3}{2} (q+n-1) \left[\widehat{\rm w}_{n-\frac{3}{2}}^{q+\frac{1}{2}},\mathcal{O}_{h,\bar{h}}\right].           
    \end{equation}

	   \subsection{$\widehat{\rm w}^p_m$ Commutator for All $m$} \label{app:allm}
	   
	   Here we show that, for fixed $p$ and arbitrary $q$ and $n \in [1-q, q-1]$,
	   if \eqref{jacobi-id} is assumed to hold for any single $m \in [1-p, p-1]$, then it also holds for all $m\in [1-p, p-1]$.
	   
	   First, note that because $\widehat{\rm w}^2_m$ generates ${\rm SL}(2, \mathbb{R})_R$ transformations (see  \eqref{eq:sl2r-w}) and the modes $\widehat{\rm w}^q_n$ transform covariantly under ${\rm SL}(2, \mathbb{R})_R$ by construction,  \eqref{jacobi-id} automatically holds for $p = 2$, $m = -1, 0, 1$, and all  $q$ and $n \in [1-q, q-1]$:\footnote{This is also straightforward to verify by brute force.} 
	    \begin{equation} \label{jacobi-id-w2}
	        \begin{split} 
	            \left[\widehat{\rm w}^2_m, \left[\widehat{\rm w}^q_n, \mathcal{O}_{h,\bh} \right]\right]
	            -  \left[\widehat{\rm w}^q_n, \left[\widehat{\rm w}^2_m, \mathcal{O}_{h,\bh} \right]\right]
	            =  \left(m (q-1)-n \right)\left[\widehat{\rm w}^{q}_{m+n},\mathcal{O}_{h,\bh}  \right].
	        \end{split}
	    \end{equation} 
	    Next, let us assume \eqref{jacobi-id} holds for a given  $p$ and $m \in [1-p, p-1]$ and all $q$ and $n \in [1-q, q-1]$.  Acting with $\widehat{\rm w}^2_{k} = \bar{L}_k$, we find 
	    \begin{equation} \label{step1-m}
	        \begin{split} 
	          \left[\widehat{\rm w}^2_k,  \left[\widehat{\rm w}^p_m, \left[\widehat{\rm w}^q_n, \mathcal{O}_{h,\bh} \right]\right]\right]
	           -  \left[\widehat{\rm w}^2_k,  \left[\widehat{\rm w}^q_n, \left[\widehat{\rm w}^p_m, \mathcal{O}_{h,\bh} \right]\right]\right]
	            =  \left(m (q-1)-n(p-1)\right) \left[\widehat{\rm w}^2_k, \left[\widehat{\rm w}^{p+q-2}_{m+n},\mathcal{O}_{h,\bh}  \right]\right].
	        \end{split}
	    \end{equation}
	    Then, we note that the first term can be manipulated as follows:
	    \begin{equation} \label{firstterm-m}
	        \begin{split}
	            & \left[\widehat{\rm w}^2_k,  \left[\widehat{\rm w}^p_m, \left[\widehat{\rm w}^q_n, \mathcal{O}_{h,\bh} \right]\right]\right]
	              = \mathcal{D}^q_n(\bh)
	                 \left[\widehat{\rm w}^2_k,  \left[\widehat{\rm w}^p_m,  \mathcal{O}_{h+2-q,\bh+2-q} \right]\right] \\
	        & \quad  = \mathcal{D}^q_n(\bh)
	                \left[\widehat{\rm w}^p_m, \left[\widehat{\rm w}^2_k, \mathcal{O}_{h+2-q,\bh+2-q} \right]\right]
	            +  \left(k (p-1)-m \right)\mathcal{D}^q_n(\bh)\left[\widehat{\rm w}^{p}_{k+m},\mathcal{O}_{h+2-q,\bh+2-q}  \right]  \\
	        & \quad   = 
	                \left[\widehat{\rm w}^p_m, \left[\widehat{\rm w}^2_k,\left[\widehat{\rm w}^q_n, \mathcal{O}_{h,\bh} \right]\right]\right]
	            +  \left(k (p-1)-m \right) \left[\widehat{\rm w}^{p}_{k+m},\left[\widehat{\rm w}^q_n, \mathcal{O}_{h,\bh} \right] \right]  \\
	       & \quad   = 
	                \left[\widehat{\rm w}^p_m,  \left[\widehat{\rm w}^q_n, \left[\widehat{\rm w}^2_k, \mathcal{O}_{h,\bh} \right]\right]\right]
	             +  \left(k (q-1)-n \right) \left[\widehat{\rm w}^p_m, \left[\widehat{\rm w}^{q}_{k+n},\mathcal{O}_{h,\bh}  \right]\right] 
	           +  \left(k (p-1)-m \right) \left[\widehat{\rm w}^{p}_{k+m},\left[\widehat{\rm w}^q_n, \mathcal{O}_{h,\bh} \right] \right], 
	        \end{split}
	    \end{equation}  
	   where  \eqref{wcharge-action-app} is used in the first and  third  lines and \eqref{jacobi-id-w2} in  the second and fourth lines.  Using \eqref{wcharge-action-app} one more time, this term can be put in the following form: 
	    \begin{equation}\label{firstterm-simp-m}
	        \begin{split}
	              \left[\widehat{\rm w}^2_k,  \left[\widehat{\rm w}^p_m, \left[\widehat{\rm w}^q_n, \mathcal{O}_{h,\bh} \right]\right]\right]
	              &  = 
	               \mathcal{D}^2_k(\bh) \left[\widehat{\rm w}^p_m,  \left[\widehat{\rm w}^q_n,   \mathcal{O}_{h,\bh} \right]\right]
	             +  \left(k (q-1)-n \right) \left[\widehat{\rm w}^p_m, \left[\widehat{\rm w}^{q}_{k+n},\mathcal{O}_{h,\bh}  \right]\right] \\& 
	             \quad\quad 
	           +  \left(k (p-1)-m \right) \left[\widehat{\rm w}^{p}_{k+m},\left[\widehat{\rm w}^q_n, \mathcal{O}_{h,\bh} \right] \right].
	        \end{split}
	    \end{equation}
	    Notice that none of the steps to arrive at \eqref{firstterm-simp-m} depended on our assumption of fixed $p$ and $m$ and arbitrary $q$ and $n$. Therefore, we can perform precisely the same steps on the second term in \eqref{step1-m} and the result is given by exchanging  $p,m \leftrightarrow q,n$ in \eqref{firstterm-simp-m}.
	    
	     Using \eqref{jacobi-id-w2} and then \eqref{wcharge-action-app} to rewrite the last term in \eqref{step1-m}, we find
	    \begin{equation}
	        \begin{split} \label{term3-m}
	            \left[\widehat{\rm w}^2_k, \left[\widehat{\rm w}^{p+q-2}_{m+n},\mathcal{O}_{h,\bh}  \right]\right]
	               &=\mathcal{D}^2_k(\bh) \left[\widehat{\rm w}^{p+q-2}_{m+n},   \mathcal{O}_{h,\bh} \right]
	            +  \left(k (p+q-3)-(m+n) \right)\left[\widehat{\rm w}^{p+q-2}_{m+n+k},\mathcal{O}_{h,\bh}  \right].
	        \end{split}
	    \end{equation}
	    Finally, using \eqref{firstterm-simp-m}
	    and \eqref{term3-m}, \eqref{step1-m} can be brought to the form 
	    \begin{equation} \label{step2-m}
	        \begin{split} 
	            \mathcal{D}^2_k(\bh) 
	           \left \{\left[\widehat{\rm w}^p_m,  \left[\widehat{\rm w}^q_n,   \mathcal{O}_{h,\bh} \right]\right]
	       -\left[\widehat{\rm w}^q_n,  \left[\widehat{\rm w}^p_m,   \mathcal{O}_{h,\bh} \right]\right]
	          - \left(m (q{-}1){-}n(p{-}1)\right)\left[\widehat{\rm w}^{p+q-2}_{m+n},   \mathcal{O}_{h,\bh} \right]
	          \right \}\\
	             {+}  \left(k (q{-}1){-}n \right) \left \{\left[\widehat{\rm w}^p_m, \left[\widehat{\rm w}^{q}_{k+n},\mathcal{O}_{h,\bh}  \right]\right]
	             {-}\left[\widehat{\rm w}^{q}_{k+n}, \left[\widehat{\rm w}^p_m,\mathcal{O}_{h,\bh}  \right]\right]
	             {-}\left(m (q{-}1){-}(n{+}k)(p{-}1)\right)\left[\widehat{\rm w}^{p+q-2}_{m+n+k},   \mathcal{O}_{h,\bh} \right] \right\} \\ 
	           {+}  \left(k (p{-}1){-}m \right) \left \{\left[\widehat{\rm w}^{p}_{k+m},\left[\widehat{\rm w}^q_n, \mathcal{O}_{h,\bh} \right] \right]{-}
	           \left[\widehat{\rm w}^q_n,\left[\widehat{\rm w}^{p}_{k+m}, \mathcal{O}_{h,\bh} \right] \right]
	           {-} \left((m{+}k) (q{-}1){-}n(p{-}1)\right)\left[\widehat{\rm w}^{p+q-2}_{m+n+k},   \mathcal{O}_{h,\bh} \right]\right\}\\
	           =0.
	        \end{split}
	    \end{equation}
	    The first and second lines vanish by the assumption that \eqref{jacobi-id} holds for all $q$ and $n$.   Then, provided that $m \neq k (p-1)$, we find 
	    \begin{equation}
	        \left[\widehat{\rm w}^{p}_{k+m},\left[\widehat{\rm w}^q_n, \mathcal{O}_{h,\bh} \right] \right]{-}
	           \left[\widehat{\rm w}^q_n,\left[\widehat{\rm w}^{p}_{k+m}, \mathcal{O}_{h,\bh} \right] \right]
	           {-} \left((m{+}k) (q{-}1){-}n(p{-}1)\right)\left[\widehat{\rm w}^{p+q-2}_{m+n+k},   \mathcal{O}_{h,\bh} \right]
	           =0.
	    \end{equation}
	    Setting $k = \pm 1$, we find that   if \eqref{jacobi-id} holds for a given $p$ and $m$, 
	    it also holds for $p$ and $m\pm1$, provided that $m \neq \pm (p-1)$.  Therefore, for a given value of $p$, if \eqref{jacobi-id} holds for any single $m$ in the range $1-p\leq m\leq p-1$, it also holds for all $m$ in  the range $1-p\leq m\leq p-1$.

	    \subsection{$\widehat{\rm w}^p_m$ Commutator for All $p$ and $m$}
	    \addtocontents{toc}{\protect\setcounter{tocdepth}{2}}
	    
	    In this subsection, we show that \eqref{jacobi-id} holds for all $p$, $q$, $m \in [1-p,p-1]$ and $n\in[1-q,q-1]$. The logic here is precisely analogous to that in the previous section so we include fewer details. 
	    
	    First, \eqref{eq:w52} together with the results from the previous subsection imply for $-\frac{3}{2} \leq m\leq \frac{3}{2}$ and all $q$ and $n\in [1-q, q-1]$ that 
	    \begin{equation} \label{jacobi-id-w5/2}
	        \begin{split}
	            \left[\widehat{\rm w}^{\frac{5}{2}}_m, \left[\widehat{\rm w}^q_n, \mathcal{O}_{h,\bh} \right]\right]
	            -  \left[\widehat{\rm w}^q_n, \left[\widehat{\rm w}^{\frac{5}{2}}_m, \mathcal{O}_{h,\bh} \right]\right]
	            =  \left(m (q-1)- \frac{3}{2}n\right)\left[\widehat{\rm w}^{q+ \frac{1}{2}}_{m+n},\mathcal{O}_{h,\bh} \right].
	        \end{split}
	    \end{equation}
	    
	    Then, assuming \eqref{jacobi-id} holds for a given $p$ and all  $m \in [1-p, p-1]$, $q$, and $n\in [1-q, q-1]$, we act with  $\widehat{\rm w}^{\frac{5}{2}}_k$ and find 
	    \begin{equation} \label{step1-p}
	        \begin{split} 
	          \left[\widehat{\rm w}^{\frac{5}{2}}_k,  \left[\widehat{\rm w}^p_m, \left[\widehat{\rm w}^q_n, \mathcal{O}_{h,\bh} \right]\right]\right]
	           -  \left[\widehat{\rm w}^{\frac{5}{2}}_k,  \left[\widehat{\rm w}^q_n, \left[\widehat{\rm w}^p_m, \mathcal{O}_{h,\bh} \right]\right]\right]
	            =  \left(m (q-1)-n(p-1)\right) \left[\widehat{\rm w}^{\frac{5}{2}}_k, \left[\widehat{\rm w}^{p+q-2}_{m+n},\mathcal{O}_{h,\bh}  \right]\right].
	        \end{split}
	    \end{equation}
	    Following a similar series of steps as in \eqref{firstterm-m} and \eqref{firstterm-simp-m}, the first term in \eqref{step1-p} can be put in the form 
	    \begin{equation} \label{firstterm-p}
	        \begin{split}
	             \left[\widehat{\rm w}^{\frac{5}{2}}_k,  \left[\widehat{\rm w}^p_m, \left[\widehat{\rm w}^q_n, \mathcal{O}_{h,\bh} \right]\right]\right]
	             & =\mathcal{D}^{\frac{5}{2} }_{k}(\bh) \left[ \widehat{\rm w}^p_m, \left[\widehat{\rm w}^q_n,  \mathcal{O}_{h-\frac{1}{2}, \bh-\frac{1}{2}} \right]\right]
				 +\left(k(q-1)- \frac{3}{2}n  \right)\left[ \widehat{\rm w}^p_m, \left[ \widehat{\rm w}^{ q+\frac{1}{2}}_{n+k}, \mathcal{O}_{h, \bh} \right]\right] \\& \quad\quad  
				+ \left(k(p-1)- \frac{3}{2}m  \right) \left[\widehat{\rm w}^{ p+\frac{1}{2}}_{m+k},  \left[ \widehat{\rm w}^q_n, \mathcal{O}_{h,\bh}\right]  \right] .
	        \end{split}
	    \end{equation}
	   Similarly, the second term in \eqref{step1-p} can be written as \eqref{firstterm-p} with $p,m \leftrightarrow q,n$.  Finally, the last term in \eqref{step1-p} can be put in the form 
	   \begin{equation}
	       \begin{split}
	           \left[\widehat{\rm w}^{\frac{5}{2}}_k, \left[\widehat{\rm w}^{p+q-2}_{m+n},\mathcal{O}_{h,\bh}  \right]\right]
	   &= \mathcal{D}^{\frac{5}{2}}_{k} (\bh)\left[  \widehat{\rm w}^{p+q-2}_{m+n},\mathcal{O}_{h-\frac{1}{2}, \bh-\frac{1}{2}} \right] 
						+ \left(k(p+q-3)- \frac{3}{2}(m+n)\right) \left[  \widehat{\rm w}^{p+q-\frac{3}{2}}_{m+n+k}, \mathcal{O}_{h, \bh}  \right].
	       \end{split}
	   \end{equation}
	   Using these results together with the assumption that  \eqref{jacobi-id} holds for a given $p$ and all $m$, $q$, and $n$,  \eqref{step1-p} implies that
	   \begin{equation}   \label{step2-p}
			\begin{split}
		\left(k(p-1)-\frac{3}{2}m  \right)\Bigg\{ &\left[\widehat{\rm w}^{ p+\frac{1}{2}}_{m+k},  \left[ \widehat{\rm w}^q_n, \mathcal{O}_{h,\bh}\right]  \right]  				 - \left[\widehat{\rm w}^q_n, \left[\widehat{\rm w}^{ p+\frac{1}{2}}_{m+k},  \mathcal{O}_{h,\bh}\right] \right] \\
			& \quad  -  \left(\left(m+k\right)(q-1)-n \left(p-\frac{1}{2}\right)\right)  \left[  \widehat{\rm w}^{p+q-\frac{3}{2}}_{m+n+k}, \mathcal{O}_{h, \bh}  \right]\Bigg \}  =0.
			\end{split}
		\end{equation} Now setting $k = \pm \frac{3}{2}$, we conclude that if \eqref{jacobi-id} holds for any given $p$ and all $m \in [1-p,p-1]$, $q$ and $n \in [1-q,q-1]$, it also holds for $p+\frac{1}{2}$, $m \pm \frac{3}{2}$ provided that $m \neq \pm (p-1)$. Then, the result from the previous subsection implies  \eqref{jacobi-id} holds for $p +\frac{1}{2}$ and all $m\in [\frac{1}{2}-p,p- \frac{1}{2}]$. 
		
		We have already shown in Subsection \ref{app:special} that \eqref{jacobi-id} holds for $(p,m) = (1,0)$ and $(p,m) = (\frac{3}{2},-\frac{1}{2})$.  Then, the results from the previous subsection imply that \eqref{jacobi-id} also holds for $(p,m) = (\frac{3}{2},\frac{1}{2})$. The results from this section further imply that \eqref{jacobi-id} holds for all $p>\frac{3}{2}$ and $m \in [1-p, p-1]$.

	\end{appendix}

\bibliography{ssope}
\bibliographystyle{utphys}

\end{document}